\documentclass{article}

\usepackage[round]{natbib}
\usepackage{amsmath,amssymb,amsthm}
\usepackage{booktabs}
\usepackage{color}
\usepackage{subfig}        
\usepackage{graphicx}      

\usepackage{algorithm}     
\usepackage{algpseudocode} 

\usepackage[colorlinks=true,citecolor=blue,urlcolor=blue,linkcolor=blue]{hyperref}
\usepackage[nameinlink]{cleveref}

\theoremstyle{plain}
\newtheorem{theorem}{Theorem}[section]
\newtheorem{lemma}{Lemma}[section]

\theoremstyle{definition}

\newtheorem{example}{Example}[section]

\theoremstyle{remark}
\newtheorem{remark}{Remark}[section]

\crefname{condition}{condition}{conditions}
\Crefname{condition}{Condition}{Conditions}

\crefname{definition}{definition}{definitions}
\Crefname{definition}{Definition}{Definitions}

\crefname{theorem}{theorem}{theorems}
\Crefname{theorem}{Theorem}{Theorems}

\crefname{lemma}{lemma}{lemmas}
\Crefname{lemma}{Lemma}{Lemmas}

\crefname{remark}{remark}{remarks}
\Crefname{remark}{Remark}{Remarks}

\crefname{example}{example}{examples}
\Crefname{example}{Example}{Examples}

\crefname{case}{case}{cases}
\Crefname{case}{Case}{Cases}

\crefname{proposition}{proposition}{propositions}
\Crefname{proposition}{Proposition}{Propositions}

\newcommand\BibTeX{{\rmfamily B\kern-.05em \textsc{i\kern-.025em b}\kern-.08em
T\kern-.1667em\lower.7ex\hbox{E}\kern-.125emX}}

\usepackage[
  letterpaper,
  top=1.3in, bottom=1.3in,
  left=1.3in, right=1.3in
]{geometry}

\usepackage{xcolor}

 \usepackage{multirow} 
 \usepackage{authblk}

 \usepackage{algorithm}
\usepackage{algpseudocode} 

\usepackage[skip=4pt]{caption}

\usepackage{float}
\usepackage{placeins}

\newtheorem{proposition}{Proposition}[section]
\newtheorem{corollary}{Corollary}[section]

\usepackage{xcolor}

\title{A Family-Level Graphical Gatekeeping Procedure with Iterative Retesting}

\author[1]{Zhiying Qiu}
\author[2]{Wenge Guo\thanks{E-mail: zhiying.qiu@sanofi.com; wenge.guo@njit.edu; sanat.sarkar@temple.edu}}
\author[3]{Sanat Sarkar}

\affil[1]{Biostatistics, Insmed, Bridgewater, NJ, USA}
\affil[2]{Department of Mathematical Sciences, New Jersey Institute of Technology, Newark, NJ, USA}
\affil[3]{Department of Statistics, Operations and Data Science, Temple University, Philadelphia, PA, USA}

\date{\today}

\begin{document}
\maketitle

\begin{abstract}
Confirmatory clinical trials often involve multiple
objectives grouped into families according to scientific
priorities. At the intersection of graphical multiple
testing and gatekeeping methods, we develop a
Bonferroni-based family-level graphical procedure with
partial-rejection-based transfers and iterative re-testing.
Families are tested sequentially within each stage,
using fixed component tests and a prespecified transition
matrix. The updating rule combines current-stage forward
contributions based on current family levels with
previous-stage backward contributions based on initial
allocations. We establish strong global familywise
error rate control at a prespecified level under arbitrary
dependence. The construction permits backward feedback
while retaining fixed component tests and transition
coefficients, distinguishing its implementation from
forward-only family-level procedures and the specified
superchain algorithm. Simulations demonstrate gains over
the corresponding No-retest procedure, while identifying
an efficiency cost relative to a hypothesis-level
graphical comparator with matched initial allocations
and edge weights. Clinical illustrations show how
the procedure implements family-level priorities and
feedback, and an extension accommodates a two-layer
structure with multiple families per layer.
\end{abstract}

\section{Introduction}\label{sec:intro}

Confirmatory clinical trials often involve multiple research
objectives organized into families according to clinical
importance and scientific priorities. Multiple testing
procedures must translate these priorities into prespecified
testing opportunities while controlling false conclusions
across all objectives. Gatekeeping and graphical methods
provide complementary approaches to this task: gatekeeping
specifies conditions for testing across families, whereas
graphs represent significance-level allocations and transfers.

Gatekeeping procedures test ordered families subject to
prespecified conditions
\citep{maurer1995multiple,bauer1998testing}.
Serial gatekeeping requires rejection of all hypotheses
in preceding families \citep{westfall2001optimally},
whereas parallel gatekeeping permits progression following
partial rejection \citep{dmitrienko2003gatekeeping}.
Tree-structured, mixture-based, and multistage procedures
accommodate more elaborate testing relationships and
level-transfer rules
\citep{dmitrienko2007tree,dmitrienko2011mixtures,
dmitrienko2011multistage,dmitrienko2016mixture,
kordzakhia2018enhanced}.
Reviews of these developments are provided by
\citet{dmitrienko2013key,alosh2014advanced,tamhane2018advances}.

Graphical multiple testing procedures make allocations
and transfers explicit. In the approach of
\citet{bretz2009graphical}, nodes represent individual
hypotheses and directed edges determine how significance
levels are recycled after rejection. Subsequent developments
extend this approach to structured and adaptive testing
settings
\citep{bretz2014graphical,maurer2013multiple,
maurer2014note,sugitani2018flexible,
robertson2020graphical,xi2024optimal,gou2024reverse}.
Graphical and gatekeeping methods overlap: graphs can
encode testing priorities and implement gatekeeping
strategies. At their intersection, family-level graphical
constructions provide a direct representation of allocations
and transfers among groups of hypotheses.

Re-testing provides a useful perspective for distinguishing
existing constructions. In hypothesis-level graphical
procedures, rejection triggers significance-level recycling,
which can allow previously unrejected hypotheses to be
tested again \citep{bretz2009graphical}.
At the family level, the Bonferroni parallel gatekeeping
procedure of \citet{guilbaud2007bonferroni} and the
$\alpha$-exhaustive multistage procedure of
\citet{dmitrienko2011multistage} permit backward testing
after complete rejection of the final family, with further
backward progression subject to complete rejection of
the family just revisited.
Algorithm~1 of \citet{maurer2014note} accommodates general
within-family procedures but permits inter-family transfer
only after complete rejection of the source family.
Superchain procedures permit partial-rejection-based
transfers and iterative re-testing through component
updates and graph reweighting
\citep{kordzakhia2013superchain}.
In contrast, the family-level graphical framework of
\citet{qiu2026graphical} accommodates general component
procedures through error-rate-function bounds, but its
Condition~1.2 restricts transfers to subsequent layers,
yielding an acyclic graph and a single forward testing pass.

This paper develops a Bonferroni-based family-level
graphical procedure with partial-rejection-based transfers
and iterative re-testing. Each node represents a family,
initial family allocations express testing priorities,
and a prespecified transition matrix determines permitted
transfers. Families are tested sequentially within each
stage. The updating rule combines current-stage forward
contributions based on current family levels with
previous-stage backward contributions based on initial
family allocations. Transfers depend on the proportions
of rejected hypotheses, allowing partial rejection in
one family to support further testing in another.
The Bonferroni component tests, original family sizes,
and transition matrix remain fixed throughout.

The contribution lies in the specification and theoretical
validation of this family-level updating scheme.
It permits backward feedback following partial rejection
without requiring component-parameter updates or graph
reweighting. We establish strong global familywise error
rate (FWER) control at level $\alpha$ under arbitrary
dependence, accounting for the successive
rejection-dependent testing levels. Here, global FWER
is the probability of rejecting at least one true null
hypothesis across all families. Relative to general
family-level frameworks, the construction supports
repeated forward and backward testing while restricting
its component tests to Bonferroni. A detailed comparison
with superchain procedures follows
Algorithm~\ref{alg:multifamily}.

The family-level formulation provides a direct way to
specify allocation priorities and feedback across families.
Strict gatekeeping restrictions must be encoded through
the chosen allocations and transfers; positive initial
allocations can permit testing without an upstream
rejection. Our numerical study compares the proposed
procedure with its No-retest version, the Bonferroni
parallel gatekeeping procedure (BR) of
\citet{guilbaud2007bonferroni}, the $\alpha$-exhaustive
multistage gatekeeping procedure (MR) of
\citet{dmitrienko2011multistage}, the superchain procedure
(SC) of \citet{kordzakhia2013superchain},
a hypothesis-level graphical Bonferroni procedure with
matched initial allocations and edge weights, and
pooled equal-weighted Holm.
The results demonstrate gains over No-retest and identify
an efficiency cost relative to the matched graphical
comparator in the studied settings.

The remainder of the paper is organized as follows.
Section~\ref{sec:prelim} introduces motivating examples,
notation, and guidance for specifying allocations and
transfers.
Section~\ref{sec:main} presents the procedures,
theoretical properties, and connections with existing
methods.
Section~\ref{sec:simulation} reports the simulation study,
and Section~\ref{sec:examples} illustrates implementation
using the clinical examples.
Section~\ref{sec:extension} extends the construction to
a two-layer structure with multiple families per layer.
Section~\ref{sec:conclusion} concludes with a discussion.
Technical proofs are provided in the Appendix.
\section{Preliminaries}\label{sec:prelim}

We introduce two motivating clinical trial structures, establish
notation, and discuss how to specify family allocations and
transfers. The examples are revisited in Section~\ref{sec:examples}
using hypothetical $p$-values rather than observed trial results.

\subsection{Motivating examples}\label{sec:motiva}

\begin{example}[Two-family problem]\label{sec:examp1}
\normalfont

This example is motivated by the EPHESUS trial
\citep{pitt2003eplerenone}, which evaluated the effects of
eplerenone on morbidity and mortality. We consider two primary
and two secondary endpoints, grouped as follows:
\begin{itemize}
\item $F_1$: all-cause mortality (P1) and the composite of
cardiovascular mortality or cardiovascular hospitalization (P2);
\item $F_2$: cardiovascular mortality (S1) and the composite of
all-cause mortality or all-cause hospitalization (S2).
\end{itemize}
For each composite endpoint, either component constitutes an event.
The corresponding null hypotheses are $H_{11},H_{12}$ for the
primary endpoints and $H_{21},H_{22}$ for the secondary endpoints.

At global level $\alpha=0.05$, assign initial family levels
$(\alpha_1,\alpha_2)=(0.04,0.01)$ and transition matrix
\[
\mathbf G=\begin{pmatrix}0&1\\1&0\end{pmatrix}.
\]
The larger allocation prioritizes $F_1$, while the positive
allocation to $F_2$ permits secondary testing without a primary
rejection. Bidirectional transfers allow primary rejections to
support secondary testing and secondary rejections to support
re-testing of the primary family. Contributions depend on the
source family's rejection fraction and the updating rules
specified in Section~\ref{sec:main}.

In the family-level graph, inspired by
\citet{bretz2009graphical}, nodes represent families and directed
edges specify transition coefficients (Figure~\ref{GRAPH_EX1}).

\begin{figure}[htbp]
\centering
\includegraphics[width=0.2\textwidth]{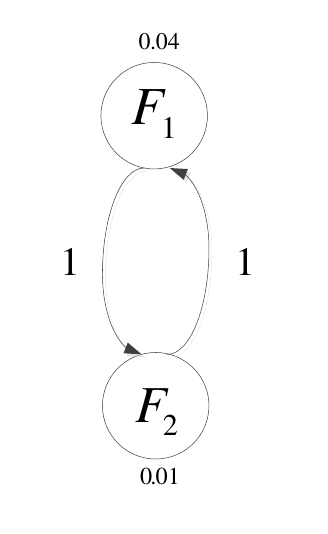}
\caption{Two-family graph with initial allocations $(0.04,0.01)$
and unit transition coefficients in both directions.}
\label{GRAPH_EX1}
\end{figure}

\end{example}

\begin{example}[Three-family problem]\label{sec:examp2}
\normalfont

We use the trial structure considered by
\citet{kordzakhia2013superchain}, comparing two treatment doses
(D1, D2) with placebo (P) in the general population (Group 1)
and two prespecified subpopulations (Groups 2 and 3).
The six hypotheses form three families:
\begin{itemize}
\item $F_1$: $H_{11}$ (D1 vs.\ P in Group 1) and
$H_{12}$ (D2 vs.\ P in Group 1);
\item $F_2$: $H_{21}$ (D1 vs.\ P in Group 2) and
$H_{22}$ (D2 vs.\ P in Group 2);
\item $F_3$: $H_{31}$ (D1 vs.\ P in Group 3) and
$H_{32}$ (D2 vs.\ P in Group 3).
\end{itemize}
At global level $\alpha=0.025$, specify
\[
(\alpha_1,\alpha_2,\alpha_3)
=(\alpha/2,\alpha/3,\alpha/6),\qquad
\mathbf G=\begin{pmatrix}
0&0.5&0.5\\0.5&0&0.5\\0.5&0.5&0
\end{pmatrix}.
\]
All families can be tested initially, with unequal levels
expressing their priorities. Each family directs one-half of its
rejection-dependent contribution to each other family.
Earlier-family rejections can increase later-family levels within
a stage, while later-family rejections can support re-testing
of earlier families in subsequent stages.
Figure~\ref{GRAPH_EX2} shows the corresponding graph.

\begin{figure}[htbp]
\centering
\includegraphics[width=0.65\textwidth]{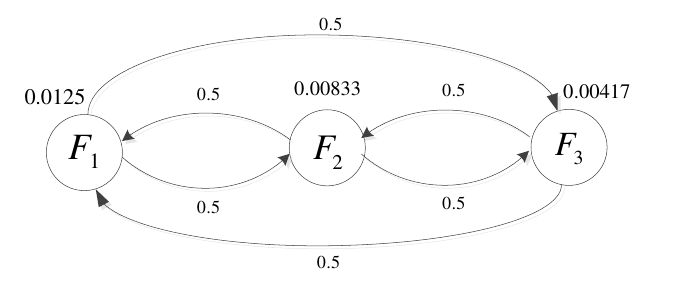}
\caption{Three-family graph with initial allocations
$(\alpha/2,\alpha/3,\alpha/6)$, $\alpha=0.025$, and transition
coefficient $0.5$ between distinct families.}
\label{GRAPH_EX2}
\end{figure}

\end{example}

\subsection{Notation and definitions}\label{sec:notation}

Let $n$ hypotheses be partitioned into $m\geq2$ ordered families
$F_i=\{H_{i1},\ldots,H_{in_i}\}$, with
$\sum_{i=1}^m n_i=n$. Write $p_{ij}$ for the $p$-value of
$H_{ij}$. For every true null, assume marginal validity:
\[
\Pr(p_{ij}\leq u)\leq u,\qquad u\in[0,1].
\]
No dependence restriction is imposed on the $p$-values.
The global FWER is the probability of rejecting at least one
true null across all families. Strong control at level $\alpha$
means that this probability is at most $\alpha$ under every
configuration of true and false nulls.

\begin{figure}[htbp]
\centering
\includegraphics[width=0.4\textwidth]{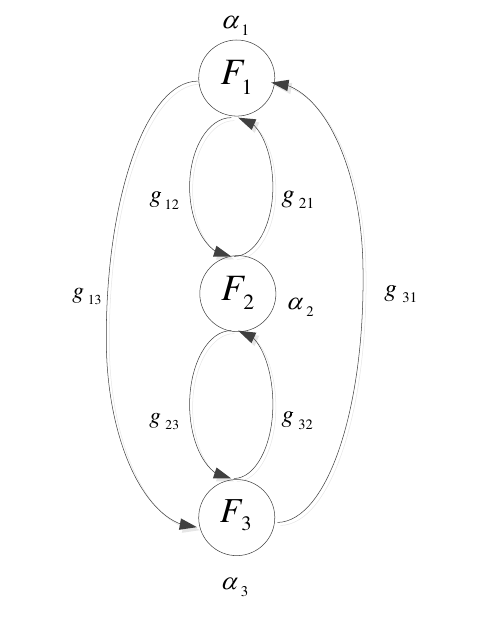}
\caption{Family-level graphical representation of allocations
and transition coefficients for three families.}
\label{THREEF_PIC}
\end{figure}

Each family receives an initial level $\alpha_i\geq0$, with
$\sum_i\alpha_i=\alpha$. The proposed procedure tests families
sequentially within each stage using Bonferroni thresholds and
updates their levels across stages. Permitted contributions are
specified by an $m\times m$ transition matrix
$\mathbf G=(g_{ij})$, satisfying
\[
g_{ij}\geq0,\qquad g_{ii}=0,\qquad
\sum_{j=1}^m g_{ij}\leq1.
\]
The coefficient $g_{ij}$ determines the fraction of the
rejection-dependent source contribution directed from $F_i$
to $F_j$; any remaining fraction is not transferred.
A family-level representation for $m=3$ appears in
Figure~\ref{THREEF_PIC}.

\begin{remark}
\normalfont
Following \citet{dmitrienko2008general}, consider Bonferroni
testing of $F_i$ at a fixed level $a$. Let $R_i$ and $A_i$
denote its rejected and unrejected sets, respectively.
Their cardinalities $|R_i|$ and $|A_i|$ count the rejected
and unrejected hypotheses, with $|R_i|+|A_i|=n_i$.
The Bonferroni error-rate-function bound for $A_i$ is
$a|A_i|/n_i$, leaving the complementary quantity
\[
a-a\frac{|A_i|}{n_i}=a\frac{|R_i|}{n_i}.
\]
This motivates rejection-fraction-based transfers.
The proposed updating rule specifies the exact forward
and backward contributions. This interpretation alone
does not establish the validity of repeated testing at
rejection-dependent levels; global FWER control is
established in Section~\ref{sec:main}.
\end{remark}

\subsection{Practical specification of allocations and transfers}
\label{sec:parameter-guidance}

Initial allocations and transition coefficients should be
prespecified to reflect scientific priorities, testing restrictions,
and study-success criteria, subject to the constraints in
Section~\ref{sec:notation}. Initial allocations determine testing
opportunities before transfers occur, with Bonferroni thresholds
$\alpha_i/n_i$. Allocating the entire level to $F_1$ makes later
testing conditional on permitted rejection-dependent contributions;
positive allocations to later families allow testing from the outset.

For gatekeeping applications, the allocations and edges must
encode the intended eligibility conditions. For example, assigning
all initial level to $F_1$ and allowing only adjacent-family
forward transfers with positive coefficients makes testing of
each subsequent family conditional on a rejection in its immediate
predecessor. This implements progression after partial rejection,
rather than requiring complete rejection. Backward edges permit
additional testing in earlier families when compatible with the
scientific restrictions.

Transition coefficients specify recipients and transfer proportions.
Equal coefficients may suit equally prioritized recipients;
unequal coefficients can express different priorities.
Among scientifically acceptable specifications, simulations can
assess power for relevant success criteria and sensitivity to
plausible signal and dependence settings. The final specification
must be fixed before examining study data. Global FWER control
and compliance with a clinical hierarchy are separate requirements.
\section{Main Results}\label{sec:main}

We present the proposed family-level graphical procedure,
first for two families and then for $m$ ordered families.
We establish strong global FWER control and describe special
cases and connections with superchain and hypothesis-level
graphical procedures.

\subsection{Two-family problem}\label{sec:twofamily}

\begin{figure}[htbp]
\centering
\includegraphics[width=0.18\textwidth]{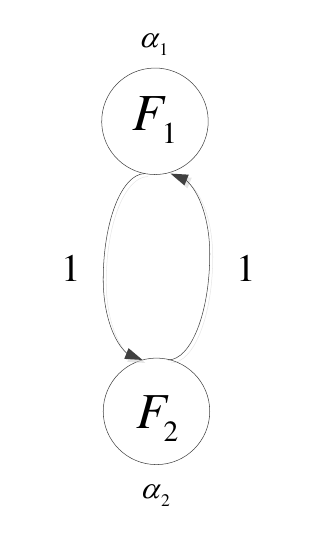}
\caption{Sequential testing and re-testing of two families.}
\label{fig:twofamily}
\end{figure}

Consider two ordered families $F_1$ and $F_2$, with initial
levels $\alpha_1,\alpha_2\geq0$ satisfying
$\alpha_1+\alpha_2=\alpha$. Let $\alpha_{i(k)}$ be the testing
level of $F_i$ in stage $k$, and let $R_{i(k)}$ be its cumulative
rejection set after that stage. We use unit transition
coefficients in both directions:
\[
\mathbf G=\begin{pmatrix}0&1\\1&0\end{pmatrix}.
\]
These coefficients specify how rejection-dependent contributions
are directed between families; the amounts are determined by
Algorithm~\ref{alg:twofamily}.
Figure~\ref{fig:twofamily} illustrates the testing sequence.

\begin{algorithm}[htbp]
\caption{Two-family Bonferroni procedure with iterative retesting}
\label{alg:twofamily}
\begin{flushleft}
\textbf{Initialization.}
Set $R_{1(0)}=R_{2(0)}=\varnothing$ and $k=1$.

\medskip
\textbf{First stage.}
Test $F_1$ by Bonferroni at level $\alpha_{1(1)}=\alpha_1$,
obtaining $R_{1(1)}$. Then test $F_2$ at level
\[
\alpha_{2(1)}=\alpha_2+\frac{|R_{1(1)}|}{n_1}\alpha_1,
\]
obtaining $R_{2(1)}$.

\medskip
\textbf{Subsequent stages.}
For $k\geq2$, first set
\begin{equation}
\alpha_{1(k)}=\alpha_1+
\frac{|R_{2(k-1)}|}{n_2}\alpha_2
\end{equation}
and test the remaining hypotheses in $F_1$. Retain earlier
rejections to obtain $R_{1(k)}$. Next set
\begin{equation}
\alpha_{2(k)}=\alpha_2+
\frac{|R_{1(k)}|}{n_1}\alpha_{1(k)}
\end{equation}
and test the remaining hypotheses in $F_2$, retaining earlier
rejections to obtain $R_{2(k)}$. The Bonferroni thresholds
are $\alpha_{i(k)}/n_i$, with the original family sizes retained.

\medskip
\textbf{Stopping rule.}
Stop if both rejection sets are unchanged from stage $k-1$,
or all hypotheses have been rejected. Otherwise increase $k$
by one and continue.
\end{flushleft}
\end{algorithm}

\newpage

\begin{proposition}[Monotonicity and finite termination]
\label{prop:twofamily}
For Algorithm~\ref{alg:twofamily}, the family levels are
nondecreasing across stages, and
\[
R_{i(1)}\subseteq R_{i(2)}\subseteq\cdots,
\qquad i=1,2.
\]
At most $n_1+n_2$ stages add rejections. Including a possible
final stage with no new rejection, the procedure terminates
within $n_1+n_2+1$ stages.
\end{proposition}

\begin{proof}[Proof sketch]
The result follows inductively. Cumulative rejection counts
cannot decrease, so the backward update cannot lower the level
of $F_1$. Its nondecreasing level and rejection count imply the
same property for the forward update of $F_2$. Bonferroni
rejection sets are therefore nested. Each stage either adds a
rejection or stops, and there are only $n_1+n_2$ hypotheses.
See Appendix~\ref{app:proof-twofamily} for details.
\end{proof}

\begin{remark}
\normalfont
The updates distinguish forward transfer from backward feedback.
Forward contributions use the current level and rejection set
of $F_1$; backward contributions use the initial allocation
of $F_2$ and its previous-stage rejection set. Levels are
recomputed from these quantities rather than increased by
repeatedly accumulating earlier transfers.
\end{remark}

\begin{remark}
\normalfont
If each family contains one hypothesis and
$\alpha_1=\alpha_2=\alpha/2$, the rejection rule coincides
with Holm's procedure for two hypotheses. A first rejection
requires a $p$-value at most $\alpha/2$ and permits testing
or re-testing the other hypothesis at level $\alpha$.
Thus, Bonferroni component tests do not imply that the complete
procedure has the rejection rule of ordinary Bonferroni testing.
\end{remark}

\subsection{Multi-family problem}\label{sec:multifamily}

For $m\geq2$ ordered families, let $\alpha_i\geq0$,
$\sum_i\alpha_i=\alpha$, and let $\mathbf G$ satisfy the
constraints in Section~\ref{sec:notation}. Define $R_{i(k)}$
as the cumulative rejection set after testing $F_i$ in stage $k$.

\begin{algorithm}[htbp]
\caption{Family-level Bonferroni procedure with iterative retesting}
\label{alg:multifamily}
\begin{flushleft}
\textbf{Initialization.}
Set $R_{i(0)}=\varnothing$ for all $i$ and $k=1$.

\medskip
\textbf{First stage.}
For $i=1,\ldots,m$ in order, compute
\[
\alpha_{i(1)}=\alpha_i+
\sum_{j<i}\frac{|R_{j(1)}|}{n_j}g_{ji}\alpha_{j(1)},
\]
and test $F_i$ by Bonferroni at threshold $\alpha_{i(1)}/n_i$.

\medskip
\textbf{Subsequent stages.}
For $k\geq2$, visit families in the same order and compute
\begin{equation}\label{eq:update_stat}
\alpha_{i(k)}=\alpha_i+
\sum_{j<i}\frac{|R_{j(k)}|}{n_j}g_{ji}\alpha_{j(k)}+
\sum_{l>i}\frac{|R_{l(k-1)}|}{n_l}g_{li}\alpha_l.
\end{equation}
Test remaining hypotheses in $F_i$ at threshold
$\alpha_{i(k)}/n_i$, retaining all earlier rejections in
$R_{i(k)}$. The denominators $n_i$ and matrix $\mathbf G$
remain fixed.

\medskip
\textbf{Stopping rule.}
Stop if every rejection set is unchanged from stage $k-1$,
or all hypotheses have been rejected. Otherwise increase $k$
by one and continue.
\end{flushleft}
\end{algorithm}

As in the two-family case, levels are nondecreasing and rejection
sets are nested. The procedure has at most $n$ stages adding
rejections and at most one final stage without a new rejection.
Equation~\eqref{eq:update_stat} propagates earlier-family
rejections forward within the current stage and incorporates
later-family rejections through backward feedback in the next
stage. Positive initial allocations to later families support
that feedback; if all their initial allocations are zero,
the backward terms vanish.

\paragraph{Relationship to superchain procedures.}
Superchain procedures also permit partial transfers and iterative
re-testing, with a broader choice of component tests.
Algorithm~3 of \citet{kordzakhia2013superchain} tests active
families simultaneously within each round and updates levels
and component error-fraction functions through $\alpha$ pumping
and power pumping. Completely rejected families are removed,
and family and connection weights are recalculated.
Algorithm~\ref{alg:multifamily} instead tests families
sequentially, retaining fixed Bonferroni components and a fixed
transition matrix. Its forward and backward contributions use
different source levels, as specified in
Equation~\eqref{eq:update_stat}. These implementation differences
provide a concrete distinction; they do not establish a general
power ordering.

\subsection{Global FWER control and special cases}\label{sec:gfwer}

\begin{theorem}\label{thm:gfwer}
Under the marginal $p$-value validity and allocation and
transition-matrix conditions in Section~\ref{sec:notation},
Algorithm~\ref{alg:multifamily} strongly controls the global
FWER at level $\alpha$ under arbitrary dependence.
\end{theorem}

The Appendix gives the proof. Fixed-level Bonferroni validity
alone is insufficient because family levels depend on rejection
outcomes. The proof bounds these levels by deterministic
quantities and applies Bonferroni inequalities, accounting for
the recursive updates and nested rejection sets.
Theorem~\ref{thm:gfwer} also applies to
Algorithm~\ref{alg:twofamily}.
Importantly, the guarantee concerns the probability of a false
rejection, not a requirement that the current family levels
sum to $\alpha$.

\begin{corollary}[Forward-only reduction]\label{cor:parallel}
If $\alpha_1=\alpha$ and $\alpha_2=\cdots=\alpha_m=0$,
all backward contributions vanish. The procedure reduces to
its first forward pass, with rejection-fraction-based transfers,
and strongly controls global FWER at level $\alpha$.
\end{corollary}

With adjacent-family forward transfers, this reduction gives
Bonferroni parallel gatekeeping: a rejection in one family
permits testing the next. More general forward connections
can permit different eligibility patterns.

\begin{corollary}[Fixed-sequence special case]\label{cor:fixedseq}
Suppose each family contains one hypothesis,
$\alpha_1=\alpha$, $\alpha_i=0$ for $i>1$, and
$g_{i,i+1}=1$ for $i=1,\ldots,m-1$, with all other entries zero.
Then Algorithm~\ref{alg:multifamily} is the classical
fixed-sequence procedure and strongly controls global FWER
at level $\alpha$.
\end{corollary}

\paragraph{Case 1: forward-only adjacent transfers.}
With positive initial allocations and
\[
g_{ij}=\begin{cases}1,&j=i+1,\\0,&\text{otherwise},\end{cases}
\]
there is no backward feedback, so subsequent stages cannot add
rejections. Each family needs only one test, and a later family
can be tested at its initial level even without an upstream
rejection. With one hypothesis per family, this is a
fallback-type construction; the fixed-sequence special case
requires the allocation in Corollary~\ref{cor:fixedseq}.

\begin{figure}[htbp]
\centering
\includegraphics[width=0.55\textwidth]{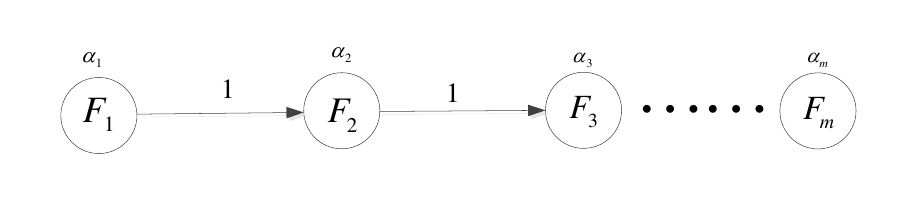}
\caption{Forward-only adjacent-family transfers.}
\label{fig:case1}
\end{figure}

\begin{figure}[htbp]
\centering
\includegraphics[width=0.55\textwidth]{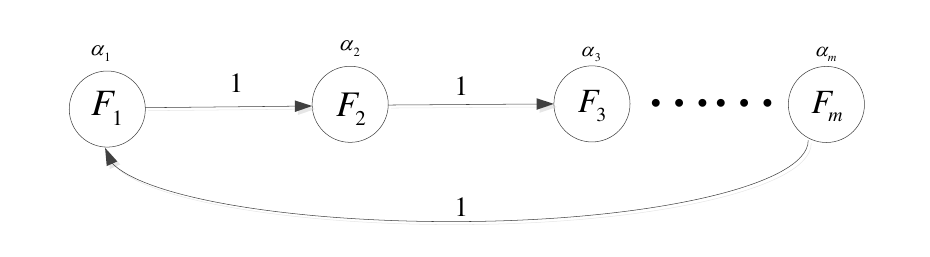}
\caption{Adjacent-family forward transfers with feedback
from the final family to the first.}
\label{fig:case2}
\end{figure}

\paragraph{Case 2: cyclic transfers.}
Add $g_{m1}=1$ to the adjacent-family forward chain, allowing
feedback from the final family to the first
(Figure~\ref{fig:case2}). For $k\geq2$,
\[
\alpha_{1(k)}=\alpha_1+
\frac{|R_{m(k-1)}|}{n_m}\alpha_m.
\]
Complete rejection of $F_m$ therefore raises the first-family
level to $\alpha_1+\alpha_m$, regardless of the current testing
level of $F_m$. A positive initial allocation $\alpha_m$ is
needed for this feedback to increase the level. With singleton
families, the construction permits re-testing within a
fallback-type allocation strategy.

\subsubsection*{Relationship to hypothesis-level graphical procedures}

The initial Bonferroni thresholds correspond to hypothesis-level
allocations $\alpha_i/n_i$. Matching these allocations and
aggregate initial transfers does not make subsequent rejection
rules identical. Retest retains original family sizes and fixed
transition coefficients; standard sequentially rejective graphs
remove rejected nodes, redistribute their current levels, and
update connections among active nodes.

For example, consider two families of size two with initial
levels $(0.04,0.01)$ and unit family transitions. The matched
hypothesis-level graph has levels $(0.02,0.02,0.005,0.005)$
and, in the order $(P_1,P_2,S_1,S_2)$, transition matrix
\[
\mathbf T=\begin{pmatrix}
0&0&1/2&1/2\\0&0&1/2&1/2\\
1/2&1/2&0&0\\1/2&1/2&0&0
\end{pmatrix}.
\]
For $p$-values $(0.010,0.029,0.001,0.019)$, Retest rejects
$P_1$ and $S_1$ in the first stage. Its next primary and
secondary thresholds are $0.0225$ and $0.01625$, so neither
remaining hypothesis is rejected and testing stops.

The matched graph rejects all four. Removing $P_1$ raises both
secondary levels to $0.015$ and gives updated weights $2/3$
from $S_1$ to $P_2$ and $1/3$ from $S_1$ to $S_2$.
Removing $S_1$ then raises their levels to $0.030$ and $0.020$,
allowing both to be rejected.
This demonstrates non-equivalence to the specified matched graph,
without excluding other graphical representations. An exact
representation would need to reproduce the complete rejection
rule and could provide an alternative error-control argument.

\paragraph{Scope of the component tests.}
The established procedures, including the two-layer extension
in Section~\ref{sec:extension}, use Bonferroni components.
Replacing them with other tests requires suitable transfer rules
and a corresponding global error-control proof. Holm is valid
under arbitrary dependence, but cannot be substituted with the
same guarantee assumed automatically. Hochberg or Hommel
components would additionally require assumptions supporting
their validity. Such extensions are not established here.
\section{Simulation Study}\label{sec:simulation}

We examine global FWER, individual and joint power, and the
benefit of iterative re-testing. Seven procedures are compared:
Retest, No-retest, the original BR and MR procedures, truncated
Holm SC, a matched hypothesis-level graphical Bonferroni
procedure (Graphical), and pooled equal-weighted Holm.

\subsection{Procedures and initial specifications}

We consider three ordered families
$F_i=\{H_{ij}:j=1,\ldots,n_i\}$ at global level $\alpha=0.05$.
Unless specified otherwise, initial family levels are
$(0.03,0.015,0.005)$, with weights $\mathbf w=(0.6,0.3,0.1)$,
and
\[
\mathbf G=\begin{pmatrix}
0&0.5&0.5\\0.5&0&0.5\\0.5&0.5&0
\end{pmatrix}.
\]
Positive initial levels permit testing without upstream
rejections; the unequal allocations express initial priorities.
The implementations below also apply to the procedures included
in Section~\ref{sec:examples}.

\paragraph{Retest and No-retest.}
Retest tests families sequentially within each stage using
Bonferroni thresholds $\alpha_{i(k)}/n_i$, retaining the original
family sizes. With cumulative rejection fractions
$r_{i(k)}=|R_{i(k)}|/n_i$, levels are recomputed as
\[
\alpha_{i(k)}=\alpha_i+
\sum_{j<i}r_{j(k)}g_{ji}\alpha_{j(k)}+
\mathbf1\{k>1\}\sum_{l>i}r_{l(k-1)}g_{li}\alpha_l.
\]
Forward contributions use current-stage levels; backward
contributions use initial allocations and previous-stage
rejection fractions. Retest stops after a stage with no new
rejections. No-retest performs only the first forward pass
with otherwise identical specifications.

\paragraph{BR and MR.}
Both published procedures start with $a_1=\alpha$ in $F_1$,
without a family transition matrix. BR follows Algorithm~$2'$
of \citet{guilbaud2007bonferroni}, using equal-weighted
Bonferroni tests in gatekeeper families and forward levels
\[
a_{i+1}=a_i|R_i|/n_i,
\]
where $R_i$ is the forward rejection set.
MR follows Section~4 of \citet{dmitrienko2011multistage},
using truncated Holm with $\gamma=0.5$ in gatekeeper families.
Its ordered $p$-values are compared with
\[
a_i\left\{\frac{\gamma}{n_i-r+1}
+\frac{1-\gamma}{n_i}\right\},\qquad r=1,\ldots,n_i,
\]
stopping at the first failure. For the unrejected set $A_i$,
\[
e_i(A_i,a_i)=
\begin{cases}
0,&A_i=\varnothing,\\
a_i\{\gamma+(1-\gamma)|A_i|/n_i\},&A_i\ne\varnothing,
\end{cases}
\qquad a_{i+1}=a_i-e_i(A_i,a_i).
\]
For both methods, a forward gate requires at least one rejection
in the preceding family, and the final family uses ordinary
Holm. Complete rejection of the final family triggers backward
Holm testing of each original family at its stored forward
level, retaining earlier rejections. Backward progression
continues only after complete rejection of the family just
revisited. Testing stops when a gate fails or the backward
pass reaches $F_1$.

\paragraph{SC.}
We implement Algorithm~3 of \citet{kordzakhia2013superchain}
with family weights $w_i=\alpha_i/\alpha$, graph $\mathbf G$,
and initial truncated Holm parameter $\gamma=0.5$.
Families are tested simultaneously within each round.
For a nonempty unrejected set $A_i$, define
$f_{i1}(A_i)=\gamma+(1-\gamma)|A_i|/n_i$.
In subsequent rounds, with remaining families $T$ and current
graph weights $w_i^*,g_{ji}^*$, use
\[
\begin{aligned}
\kappa_{ik}&=w_i^*+\sum_{j\in T\setminus\{i\}}
w_jg_{ji}^*\{1-f_{j1}(A_{j,k-1})\},\\
a_{ik}&=\alpha\kappa_{ik},\qquad
\gamma_{ik}=1-\frac{w_i}{\kappa_{ik}}(1-\gamma).
\end{aligned}
\]
For $q_i$ remaining hypotheses, the thresholds are
\[
a_{ik}\left\{\frac{\gamma_{ik}}{q_i-r+1}
+\frac{1-\gamma_{ik}}{n_i}\right\},\qquad r=1,\ldots,q_i.
\]
Completely rejected families are removed using the published
graph-reweighting rule; the last remaining family uses ordinary
Holm. SC stops when no new rejections occur or all hypotheses
are rejected.

\paragraph{Graphical.}
Following \citet{bretz2009graphical}, each node $u=(i,j)$
receives initial level $a_u=\alpha_i/n_i$, with initial edges
\[
h_{(i,j),(l,s)}=
\begin{cases}g_{il}/n_l,&l\ne i,\\0,&l=i.\end{cases}
\]
Thus, for sizes $(2,2,2)$, local levels are $0.015$, $0.0075$,
and $0.0025$, and cross-family edges have weight $0.25$.
An active node is rejected when its $p$-value does not exceed
its local level. Among eligible nodes, we select the smallest
$p$-value-to-level ratio, breaking ties by index. Removal of
node $s$ gives
\[
a_v'=a_v+a_sh_{sv},\qquad
h_{uv}'=\frac{h_{uv}+h_{us}h_{sv}}{1-h_{us}h_{su}},
\quad u\ne v,\quad u,v\ne s.
\]
Updates use the pre-removal graph, with zero diagonals and
zero outgoing rows when the denominator is zero. Testing stops
when no node is eligible. The cross-family restriction applies
only initially: graph updates can induce within-family edges.

\paragraph{Holm and comparison scope.}
Pooled Holm tests all $N=\sum_i n_i$ hypotheses jointly at
level $\alpha$. Retest versus No-retest isolates iteration.
SC shares Retest's initial family allocations and graph, while
Graphical matches its initial hypothesis levels and aggregate
transfer coefficients. BR and MR initially allocate the full
level to $F_1$, and pooled Holm assigns equal hypothesis weights.
These comparisons therefore assess complete testing strategies.

\subsection{Data generation and configurations}

We generate test statistics directly:
\[
\mathbf Z\sim N_N(\boldsymbol\mu,\boldsymbol\Sigma),
\qquad p_{ij}=1-\Phi(Z_{ij}),
\]
with unit marginal variances, zero means for true nulls, and
positive noncentralities for false nulls. No separate
subject-level sample size is specified; noncentralities reflect
the combined effects of sample size, effect size, and variability.
Within- and between-family correlations are $\rho_W$ and
$\rho_B$, represented by
\[
Z_{ij}=\mu_{ij}+\sqrt{\rho_B}U+
\sqrt{\rho_W-\rho_B}U_i+
\sqrt{1-\rho_W}\varepsilon_{ij},
\]
with independent standard normal latent variables and
$0\leq\rho_B\leq\rho_W<1$. We consider independence,
equicorrelations $0.3$ and $0.6$, and structured correlations
$(\rho_W,\rho_B)=(0.3,0.1)$ and $(0.6,0.2)$.
All methods receive the same $p$-values in each replication.

The 152 configurations form seven focused blocks rather than
a full factorial design. False nulls occupy the first specified
indices, without imposing within-family testing priorities.
\begin{enumerate}
\item \textbf{Baseline (20).} Sizes $(2,2,2)$ and false-null
counts $(0,0,0)$, $(1,0,0)$, $(1,1,0)$, $(2,1,0)$,
$(2,2,0)$, $(0,0,1)$, $(1,1,1)$, and $(2,2,2)$,
with common noncentrality $2.2$. Two further $(1,1,1)$
settings use family-specific noncentralities $(1,1.5,2.2)$
and $(2.2,1.5,1)$. Each uses independence and equicorrelation
$0.3$.
\item \textbf{Family size and sparsity (60).} Sizes $(4,4,4)$,
$(8,8,8)$, and $(2,4,8)$, with signals in $F_1$ only,
$F_3$ only, or all families. Affected families have target
false-null proportions $0.25$, $0.5$, or $1$, common
noncentrality $2.2$, and counts rounded to the nearest integer
with a minimum of one. For $n_i=2$, both the $0.25$ and
$0.5$ targets give one false null. Include the global null
for each size; use independence and correlation $(0.6,0.2)$.
\item \textbf{Signal strength (36).} Sizes $(4,4,4)$,
false-null counts $(2,0,0)$, $(0,0,2)$, or $(2,2,2)$,
and common noncentralities $0.2$, $0.3$, $1$, $1.5$, $2.2$,
or $3$, under independence and correlation $(0.3,0.1)$.
\item \textbf{Heterogeneity (6).} Sizes $(4,4,4)$ and two
false nulls per family, with noncentralities $(1,2.2)$ within
each family or family-specific values $(1,1.5,2.2)$ and
$(2.2,1.5,1)$; use independence and correlation $(0.3,0.1)$.
\item \textbf{Dependence (15).} Sizes $(4,4,4)$, false-null
counts $(0,0,0)$, $(0,0,2)$, or $(2,2,2)$, noncentrality
$2.2$, and all five covariance settings.
\item \textbf{Allocation (9).} Sizes $(4,4,4)$, counts
$(2,0,0)$, $(0,0,2)$, or $(2,2,2)$, noncentrality $2.2$,
and correlation $(0.6,0.2)$. Retest, No-retest, SC, and
Graphical use the default weights, equal family weights,
or reversed weights $(0.1,0.3,0.6)$; BR, MR, and Holm
retain their own rules.
\item \textbf{Transfer structure (6).} Sizes $(4,4,4)$,
counts $(0,0,2)$ or $(2,2,2)$, noncentrality $2.2$, and
correlation $(0.3,0.1)$. Compare the default graph with
\[
\mathbf G_{\mathrm{forward}}=
\begin{pmatrix}0&0.5&0.5\\0&0&1\\0&0&0\end{pmatrix},
\qquad
\mathbf G_{\mathrm{asym}}=
\begin{pmatrix}0&0.8&0.2\\0.2&0&0.8\\0.8&0.2&0\end{pmatrix}.
\]
\end{enumerate}
Some blocks repeat factor combinations with independent seeds;
the 152 runs are therefore not all distinct scientific settings.

\subsection{Performance measures and precision}

For replication $b$, let $V_b$ and $S_b$ count rejected true
and false nulls, and let $m_1$ denote the number of false nulls.
We estimate FWER by $B^{-1}\sum_b\mathbf1\{V_b>0\}$,
average power by $B^{-1}\sum_b S_b/m_1$, and average rejection
count by $B^{-1}\sum_b(V_b+S_b)$. Family-specific power averages
false-null rejection indicators within each family. Individual
rejection probabilities measure power for false nulls and
marginal Type~I error for true nulls.
For a nonempty target set $T$ of false nulls, joint power is
\[
\widehat\pi_T=B^{-1}\sum_b
\mathbf1\{\text{all hypotheses in }T\text{ are rejected}\}.
\]
Targets include the false nulls in each family, in $F_1\cup F_2$
when both contain signals, and across all families. Empty-target
power is not applicable; FWER is zero when all nulls are false.

Each configuration initially uses $B=10{,}000$ replications.
For a replication-level measure $Y_b$, the Monte Carlo standard
error (MCSE) is $s_Y/\sqrt B$. For probabilities, this is
approximately $\sqrt{\widehat p(1-\widehat p)/B}$: $0.0022$
at $\widehat p=0.05$ and at most $0.005$.
Average-power MCSEs use the variability of $S_b/m_1$; paired
contrasts use within-replication differences between methods.
A configuration with initial FWER more than two MCSEs above
$0.05$ receives an independent $100{,}000$-replication check.
Both estimates are reported below for the one triggered setting.
Screening seeds are $20261003+c$ for configuration index $c$;
precision-check seeds add $10^6$.

\subsection{Results and discussion}

Tables~\ref{tab:sim_indep}--\ref{tab:sim_individual_joint}
present representative average, individual, and joint power
estimates with MCSEs. Configuration IDs identify simulation runs.

\begin{table}[htbp]
\centering
\scriptsize
\caption{Average power in selected baseline configurations with family sizes $(2,2,2)$, common noncentrality $2.2$, and independence. The first column gives the false-null counts; IDs identify simulation runs. MCSEs are in parentheses.}
\label{tab:sim_indep}
\resizebox{\textwidth}{!}{%
\begin{tabular}{llccccccc}
\hline
False-null counts & ID & Retest & No-retest & BR & MR & SC & Graphical & Holm \\
\hline
$(1,0,0)$ & C002 & 0.511 (0.0050) & 0.510 (0.0050) & 0.596 (0.0049) & 0.597 (0.0049) & 0.511 (0.0050) & 0.511 (0.0050) & 0.423 (0.0049) \\
$(1,1,0)$ & C003 & 0.475 (0.0036) & 0.471 (0.0036) & 0.439 (0.0041) & 0.411 (0.0039) & 0.468 (0.0036) & 0.476 (0.0036) & 0.433 (0.0036) \\
$(2,1,0)$ & C004 & 0.503 (0.0031) & 0.498 (0.0030) & 0.544 (0.0033) & 0.560 (0.0034) & 0.520 (0.0031) & 0.505 (0.0031) & 0.450 (0.0031) \\
$(2,2,0)$ & C005 & 0.503 (0.0027) & 0.492 (0.0026) & 0.526 (0.0031) & 0.543 (0.0034) & 0.531 (0.0029) & 0.508 (0.0027) & 0.468 (0.0028) \\
$(0,0,1)$ & C006 & 0.273 (0.0045) & 0.273 (0.0045) & 0.001 (0.0002) & 0.000 (0.0002) & 0.270 (0.0044) & 0.273 (0.0045) & 0.427 (0.0049) \\
$(1,1,1)$ & C007 & 0.443 (0.0031) & 0.439 (0.0031) & 0.337 (0.0034) & 0.295 (0.0029) & 0.423 (0.0030) & 0.445 (0.0031) & 0.453 (0.0030) \\
$(2,2,2)$ & C010 & 0.479 (0.0024) & 0.463 (0.0023) & 0.464 (0.0032) & 0.448 (0.0032) & 0.509 (0.0027) & 0.510 (0.0027) & 0.513 (0.0026) \\
\hline
\end{tabular}}
\end{table}

\begin{table}[htbp]
\centering
\scriptsize
\setlength{\tabcolsep}{3pt}
\caption{Average power across signal-strength, heterogeneity, and dependence settings. All rows use family sizes $(4,4,4)$, two false nulls per family, default family allocations, and the symmetric graph. In Panel B, ``Within-family'' assigns noncentralities $(1,2.2)$ to the two false nulls in each family; the other profiles give common noncentralities within the respective families. IDs identify simulation runs; MCSEs are in parentheses.}
\label{tab:sim_dep}
\resizebox{\textwidth}{!}{%
\begin{tabular}{llccccccc}
\hline
Setting & ID & Retest & No-retest & BR & MR & SC & Graphical & Holm \\
\hline
\multicolumn{9}{l}{\textit{A. Common noncentrality; independence}} \\
$0.2$ & C083 & 0.008 (0.0004) & 0.008 (0.0004) & 0.008 (0.0004) & 0.008 (0.0004) & 0.008 (0.0004) & 0.008 (0.0004) & 0.008 (0.0004) \\
$0.3$ & C086 & 0.010 (0.0004) & 0.010 (0.0004) & 0.009 (0.0004) & 0.009 (0.0004) & 0.010 (0.0004) & 0.010 (0.0004) & 0.011 (0.0004) \\
$1.0$ & C089 & 0.052 (0.0009) & 0.052 (0.0009) & 0.040 (0.0008) & 0.040 (0.0008) & 0.051 (0.0009) & 0.052 (0.0009) & 0.053 (0.0009) \\
$1.5$ & C092 & 0.125 (0.0014) & 0.124 (0.0014) & 0.093 (0.0013) & 0.088 (0.0012) & 0.123 (0.0014) & 0.125 (0.0014) & 0.131 (0.0014) \\
$2.2$ & C095 & 0.344 (0.0021) & 0.339 (0.0020) & 0.272 (0.0023) & 0.242 (0.0020) & 0.336 (0.0020) & 0.346 (0.0021) & 0.354 (0.0021) \\
$3.0$ & C098 & 0.676 (0.0020) & 0.668 (0.0020) & 0.627 (0.0026) & 0.554 (0.0025) & 0.661 (0.0020) & 0.683 (0.0021) & 0.687 (0.0020) \\
\hline
\multicolumn{9}{l}{\textit{B. Heterogeneous noncentralities; independence}} \\
Within-family & C099 & 0.188 (0.0016) & 0.187 (0.0016) & 0.135 (0.0015) & 0.125 (0.0014) & 0.184 (0.0015) & 0.188 (0.0016) & 0.195 (0.0016) \\
$(1,1.5,2.2)$ & C100 & 0.141 (0.0015) & 0.140 (0.0014) & 0.047 (0.0011) & 0.043 (0.0009) & 0.139 (0.0014) & 0.141 (0.0015) & 0.176 (0.0015) \\
$(2.2,1.5,1)$ & C101 & 0.189 (0.0015) & 0.188 (0.0015) & 0.190 (0.0015) & 0.185 (0.0014) & 0.188 (0.0015) & 0.190 (0.0015) & 0.173 (0.0015) \\
\hline
\multicolumn{9}{l}{\textit{C. Correlations $(\rho_W,\rho_B)$; common noncentrality $2.2$}} \\
$(0,0)$ & C125 & 0.345 (0.0021) & 0.340 (0.0021) & 0.272 (0.0023) & 0.242 (0.0020) & 0.337 (0.0020) & 0.347 (0.0021) & 0.355 (0.0021) \\
$(0.3,0.3)$ & C128 & 0.342 (0.0029) & 0.337 (0.0028) & 0.299 (0.0030) & 0.265 (0.0026) & 0.334 (0.0028) & 0.346 (0.0029) & 0.352 (0.0029) \\
$(0.6,0.6)$ & C131 & 0.352 (0.0035) & 0.347 (0.0035) & 0.332 (0.0035) & 0.292 (0.0031) & 0.343 (0.0034) & 0.357 (0.0036) & 0.359 (0.0036) \\
$(0.3,0.1)$ & C134 & 0.338 (0.0025) & 0.333 (0.0024) & 0.270 (0.0026) & 0.242 (0.0023) & 0.331 (0.0024) & 0.340 (0.0025) & 0.349 (0.0025) \\
$(0.6,0.2)$ & C137 & 0.349 (0.0028) & 0.344 (0.0028) & 0.289 (0.0030) & 0.260 (0.0026) & 0.342 (0.0028) & 0.352 (0.0029) & 0.359 (0.0029) \\
\hline
\end{tabular}}
\end{table}

\paragraph{FWER and the effect of iteration.}
The largest final FWER estimate is $0.0515$ (MCSE $0.0022$),
for BR and MR under the baseline global null. Retest's maximum
is $0.0507$, and no final estimate exceeds $0.05$ by more than
two MCSEs. In C063, with sizes $(2,4,8)$, counts $(0,0,2)$,
noncentrality $2.2$, and independence, Retest's initial estimate
$0.0552$ (MCSE $0.0023$) triggered the precision check.
The independent estimate was $0.04749$ (MCSE $0.00067$).
Simulation findings are consistent with the theoretical FWER
result; they do not establish control beyond the studied settings.

Retest contains No-retest's rejection set in every replication,
with average-power gains from zero to $0.0304$. The largest
gain occurs for sizes $(4,4,4)$, two false nulls per family,
noncentrality $2.2$, correlations $(0.6,0.2)$, and weights
$(0.1,0.3,0.6)$: power rises from $0.309$ to $0.339$
(paired gain MCSE $0.0008$). Gains are often negligible for
sparse or weak signals. Retest and No-retest coincide under
the forward-only graph.

\paragraph{Comparisons with other procedures.}
BR and MR favor signals in their initially prioritized family.
Under independence with counts $(1,0,0)$, their powers are
$0.596$ and $0.597$, versus $0.511$ for Retest. With counts
$(0,0,1)$, their earlier gates restrict access to $F_3$:
power is below $0.001$, versus $0.273$ for Retest and $0.427$
for pooled Holm. These differences reflect allocation and
eligibility rules as well as component tests and recycling.

Graphical contains Retest's rejection set in every replication,
including the precision check. Its largest average-power
advantage is $0.0354$ (paired MCSE $0.0008$), for sizes
$(2,4,8)$, all false nulls, noncentrality $2.2$, and correlations
$(0.6,0.2)$: power is $0.371$ versus $0.335$ for Retest.
Joint power is $0.096$ versus $0.016$. This identifies an
efficiency cost relative to the matched graph in the studied
settings; empirical containment is not a general dominance theorem.

\begin{table}[htbp]
\centering
\scriptsize
\caption{Average and joint power for distributed signals with noncentrality $2.2$ and correlations $(\rho_W,\rho_B)=(0.6,0.2)$. The first three configurations have false-null counts $(2,2,2)$, $(4,4,4)$, and $(1,2,4)$, respectively; all hypotheses are false in C080. Joint power requires rejection of all false nulls. MCSEs are in parentheses.}
\label{tab:sim_sizes_joint}
\resizebox{\textwidth}{!}{%
\begin{tabular}{llccccccc}
\hline
Sizes and ID & Measure & Retest & No-retest & BR & MR & SC & Graphical & Holm \\
\hline
$(4,4,4)$ C037 & Average power & 0.344 (0.0028) & 0.339 (0.0028) & 0.281 (0.0030) & 0.253 (0.0026) & 0.336 (0.0028) & 0.347 (0.0029) & 0.355 (0.0029) \\
$(4,4,4)$ C037 & Joint power & 0.034 (0.0018) & 0.029 (0.0017) & 0.036 (0.0019) & 0.018 (0.0013) & 0.029 (0.0017) & 0.040 (0.0020) & 0.040 (0.0020) \\
$(8,8,8)$ C057 & Average power & 0.259 (0.0024) & 0.256 (0.0023) & 0.210 (0.0024) & 0.191 (0.0021) & 0.259 (0.0023) & 0.262 (0.0024) & 0.271 (0.0024) \\
$(8,8,8)$ C057 & Joint power & 0.005 (0.0007) & 0.004 (0.0007) & 0.005 (0.0007) & 0.002 (0.0005) & 0.005 (0.0007) & 0.006 (0.0008) & 0.006 (0.0008) \\
$(2,4,8)$ C077 & Average power & 0.300 (0.0027) & 0.296 (0.0026) & 0.222 (0.0027) & 0.181 (0.0022) & 0.287 (0.0026) & 0.303 (0.0027) & 0.339 (0.0028) \\
$(2,4,8)$ C077 & Joint power & 0.024 (0.0015) & 0.022 (0.0015) & 0.028 (0.0017) & 0.014 (0.0012) & 0.022 (0.0015) & 0.028 (0.0017) & 0.033 (0.0018) \\
$(2,4,8)$ C080 & Average power & 0.335 (0.0027) & 0.324 (0.0026) & 0.330 (0.0034) & 0.326 (0.0034) & 0.381 (0.0032) & 0.371 (0.0032) & 0.393 (0.0032) \\
$(2,4,8)$ C080 & Joint power & 0.016 (0.0013) & 0.013 (0.0011) & 0.097 (0.0030) & 0.102 (0.0030) & 0.097 (0.0030) & 0.096 (0.0029) & 0.093 (0.0029) \\
\hline
\end{tabular}}
\end{table}

\paragraph{Signals, family sizes, and dependence.}
Table~\ref{tab:sim_dep} compares all seven procedures while
varying signal strength, heterogeneity, and dependence.
At noncentralities $0.2$ and $0.3$, power is low for every
method and iteration produces little gain. With stronger
distributed signals, Retest gains over No-retest and exceeds
BR and MR in the displayed settings, while Graphical and
pooled Holm have higher power. These comparisons reflect
the procedures' different initial allocations and testing rules.

Heterogeneity changes the comparisons. When family
noncentralities are $(1,1.5,2.2)$, Retest's power is $0.141$,
versus $0.047$ for BR and $0.176$ for Holm. Reversing the
profile gives $0.189$, $0.190$, and $0.173$, respectively,
illustrating how signal placement interacts with initial
priority and forward gates. Across the five correlation
settings, Retest's power ranges from $0.338$ to $0.352$,
whereas BR's ranges from $0.270$ to $0.332$; dependence
therefore affects the methods differently. Table~\ref{tab:sim_sizes_joint}
shows the effect of larger and unequal families: with half
the hypotheses false and correlations $(0.6,0.2)$, Retest's
power decreases from $0.344$ for sizes $(4,4,4)$ to $0.259$
for $(8,8,8)$ at fixed noncentrality $2.2$.

\paragraph{Individual and joint objectives.}
In C037, Retest has higher average power than SC ($0.344$
versus $0.336$), but lower joint power for the two false nulls
in $F_1$ ($0.273$ versus $0.288$).
Joint power for all six false nulls is $0.034$ for Retest,
$0.029$ for No-retest, and $0.040$ for Graphical.
Thus, comparisons depend on the specified study-success
criterion as well as average power.

\begin{table}[htbp]
\centering
\scriptsize
\caption{Individual and joint power in C037: family sizes $(4,4,4)$, two false nulls per family with noncentrality $2.2$, and correlations $(\rho_W,\rho_B)=(0.6,0.2)$. Joint targets include only the false nulls in the indicated families. MCSEs are in parentheses.}
\label{tab:sim_individual_joint}
\resizebox{\textwidth}{!}{%
\begin{tabular}{llccccccc}
\hline
Target & Measure & Retest & No-retest & BR & MR & SC & Graphical & Holm \\
\hline
$H_{11}$ & Individual power & 0.410 (0.0049) & 0.403 (0.0049) & 0.480 (0.0050) & 0.492 (0.0050) & 0.414 (0.0049) & 0.413 (0.0049) & 0.353 (0.0048) \\
$H_{12}$ & Individual power & 0.416 (0.0049) & 0.407 (0.0049) & 0.485 (0.0050) & 0.497 (0.0050) & 0.422 (0.0049) & 0.419 (0.0049) & 0.354 (0.0048) \\
$H_{21}$ & Individual power & 0.344 (0.0048) & 0.341 (0.0047) & 0.242 (0.0043) & 0.197 (0.0040) & 0.340 (0.0047) & 0.348 (0.0048) & 0.355 (0.0048) \\
$H_{22}$ & Individual power & 0.341 (0.0047) & 0.338 (0.0047) & 0.246 (0.0043) & 0.203 (0.0040) & 0.337 (0.0047) & 0.344 (0.0048) & 0.350 (0.0048) \\
$H_{31}$ & Individual power & 0.278 (0.0045) & 0.275 (0.0045) & 0.121 (0.0033) & 0.065 (0.0025) & 0.254 (0.0044) & 0.281 (0.0045) & 0.359 (0.0048) \\
$H_{32}$ & Individual power & 0.273 (0.0045) & 0.269 (0.0044) & 0.115 (0.0032) & 0.063 (0.0024) & 0.248 (0.0043) & 0.276 (0.0045) & 0.360 (0.0048) \\
$F_1$ & Joint power & 0.273 (0.0045) & 0.265 (0.0044) & 0.339 (0.0047) & 0.363 (0.0048) & 0.288 (0.0045) & 0.278 (0.0045) & 0.227 (0.0042) \\
$F_2$ & Joint power & 0.208 (0.0041) & 0.204 (0.0040) & 0.154 (0.0036) & 0.122 (0.0033) & 0.211 (0.0041) & 0.213 (0.0041) & 0.220 (0.0041) \\
$F_3$ & Joint power & 0.163 (0.0037) & 0.159 (0.0037) & 0.078 (0.0027) & 0.036 (0.0019) & 0.151 (0.0036) & 0.167 (0.0037) & 0.237 (0.0043) \\
$F_1\cup F_2$ & Joint power & 0.091 (0.0029) & 0.083 (0.0028) & 0.107 (0.0031) & 0.091 (0.0029) & 0.093 (0.0029) & 0.098 (0.0030) & 0.082 (0.0027) \\
All families & Joint power & 0.034 (0.0018) & 0.029 (0.0017) & 0.036 (0.0019) & 0.018 (0.0013) & 0.029 (0.0017) & 0.040 (0.0020) & 0.040 (0.0020) \\
\hline
\end{tabular}}
\end{table}
\section{Clinical trial examples}\label{sec:examples}

We illustrate the clinical trial structures introduced in
Section~\ref{sec:prelim} using hypothetical $p$-values,
not observed trial results. We compare Retest, No-retest,
original BR and MR, truncated Holm SC, and pooled Holm,
using the implementations in Section~\ref{sec:simulation}.
Retest, No-retest, and SC share the stated family allocations;
BR and MR start with the full global level in $F_1$.
MR and SC use initial truncation parameter $\gamma=0.5$.
Only Retest versus No-retest isolates iterative re-testing;
the other comparisons concern complete testing strategies.

\subsection{Two-family problem}

For the EPHESUS-motivated structure in
Example~\ref{sec:examp1}, let
$F_1=\{H_{11},H_{12}\}$ and $F_2=\{H_{21},H_{22}\}$.
Set $\alpha=0.05$, family allocations $(0.04,0.01)$ for
Retest, No-retest, and SC, and unit transition coefficients
in both directions (Figure~\ref{GRAPH_EX1}).
Table~\ref{EXAMPLE1} gives two hypothetical configurations
and their final decisions.

\subsubsection{Configuration A: feedback to the primary family}

\paragraph{Retest.}
In Stage~1, $F_1$ has threshold $0.02$, rejecting only $H_{11}$.
Then $\alpha_{2(1)}=0.01+0.04/2=0.03$, with threshold
$0.015$, rejecting only $H_{21}$. No-retest ends here.
In Stage~2,
\[
\alpha_{1(2)}=0.04+0.01/2=0.045,\qquad
\alpha_{2(2)}=0.01+0.045=0.055.
\]
The primary threshold $0.0225$ rejects $H_{12}$ because
$p_{12}=0.0220$. The secondary threshold $0.0275$ does not
reject $H_{22}$, whose $p$-value is $0.0600$.
The next stage gives no new rejection, leaving
$\{H_{11},H_{12},H_{21}\}$.

\paragraph{Other procedures.}
BR rejects both primary hypotheses at threshold $0.025$;
MR rejects them at successive thresholds $0.025$ and $0.0375$.
Both then apply Holm in $F_2$ at level $0.05$, rejecting only
$H_{21}$. SC rejects both primaries at initial thresholds
$0.02$ and $0.03$, then applies Holm to the remaining family
at level $0.05$, also rejecting only $H_{21}$.
Pooled Holm reaches the same rejection set since
\[
0.0084\leq0.05/4,\qquad0.0121\leq0.05/3,
\qquad0.0220\leq0.05/2,
\]
while $0.0600>0.05$.

\subsubsection{Configuration B: successive secondary updates}

\paragraph{Retest.}
Stage~1 again rejects $H_{11}$ and $H_{21}$ at thresholds
$0.02$ and $0.015$. In Stage~2,
\[
\alpha_{1(2)}=0.045,\qquad
\alpha_{2(2)}=0.01+0.045/2=0.0325.
\]
The primary rejection set is unchanged, but the secondary
threshold $0.01625$ rejects $H_{22}$ because $p_{22}=0.0162$.
In Stage~3, $\alpha_{1(3)}=0.05$ and
$\alpha_{2(3)}=0.035$. Since $p_{12}=0.0400>0.025$,
no new rejections occur. Retest ends with
$\{H_{11},H_{21},H_{22}\}$, while No-retest rejects only
$H_{11}$ and $H_{21}$.

\paragraph{BR and MR.}
BR rejects only $H_{11}$ in $F_1$, transferring level $0.025$
to $F_2$. Its first Holm comparison fails because
$0.0140>0.025/2=0.0125$.
MR also rejects only $H_{11}$, since $0.0400>0.0375$.
The unrejected-set bound is
$e_1=0.05(0.5+0.5/2)=0.0375$, leaving secondary level
$0.0125$. Its first Holm comparison fails because
$0.0140>0.00625$. Neither method activates backward testing.

\paragraph{SC and pooled Holm.}
SC initially rejects only $H_{11}$ at primary thresholds
$0.02$ and $0.03$; $F_2$ makes no rejection at level $0.01$.
In the next round, its secondary level and truncation parameter are
\[
a_2=0.01+0.04(1-0.75)=0.02,\qquad
\gamma_2=1-\frac{0.2(1-0.5)}{0.02/0.05}=0.75.
\]
The first secondary threshold is $0.01<0.0140$, so no new
rejection occurs. The remaining primary hypothesis also fails its threshold
$0.04(0.5+0.5/2)=0.03$. SC ends with only $H_{11}$.
Pooled Holm rejects none, since its smallest $p$-value,
$0.0140$, exceeds $0.05/4=0.0125$.

\begin{table}[htbp]
\centering
\small
\caption{Final rejection decisions for two hypothetical
configurations motivated by the EPHESUS trial structure.
Retest, No-retest, and SC use initial family allocations
$(0.04,0.01)$; original BR and MR use $(0.05,0)$.
Pooled Holm tests all four
hypotheses at global level $\alpha=0.05$.
S = significant; NS = not significant.}
\label{EXAMPLE1}
\begin{tabular}{llccccccc}
\hline
Family & Hypothesis & $p$-value & Retest & No-retest
& BR & MR & SC & Holm\\
\hline
\multicolumn{9}{l}{\textbf{Configuration A}}\\
$F_1$ & $H_{11}$ & 0.0121 & S  & S  & S  & S  & S  & S\\
      & $H_{12}$ & 0.0220 & S  & NS & S  & S  & S  & S\\
$F_2$ & $H_{21}$ & 0.0084 & S  & S  & S  & S  & S  & S\\
      & $H_{22}$ & 0.0600 & NS & NS & NS & NS & NS & NS\\
\multicolumn{3}{l}{Number rejected}
& 3 & 2 & 3 & 3 & 3 & 3\\
\hline
\multicolumn{9}{l}{\textbf{Configuration B}}\\
$F_1$ & $H_{11}$ & 0.0190 & S  & S  & S  & S  & S  & NS\\
      & $H_{12}$ & 0.0400 & NS & NS & NS & NS & NS & NS\\
$F_2$ & $H_{21}$ & 0.0140 & S  & S  & NS & NS & NS & NS\\
      & $H_{22}$ & 0.0162 & S  & NS & NS & NS & NS & NS\\
\multicolumn{3}{l}{Number rejected}
& 3 & 2 & 1 & 1 & 1 & 0\\
\hline
\end{tabular}
\end{table}

\paragraph{Interpretation.}
Configuration A illustrates an additional primary rejection
through partial secondary feedback; Configuration B illustrates
an additional secondary rejection through a backward update
followed by a forward update. Both yield one more rejection
than No-retest. The differing outcomes for BR, MR, SC, and Holm
also reflect their allocation and component-testing rules,
rather than a general power ordering.

\subsection{Three-family problem}

For the dose-response structure in Example~\ref{sec:examp2},
set $\alpha=0.025$ and initial allocations
\[
(\alpha_1,\alpha_2,\alpha_3)
=(\alpha/2,\alpha/3,\alpha/6)
\]
for Retest, No-retest, and SC. The graph has off-diagonal
coefficients $0.5$ (Figure~\ref{GRAPH_EX2}).
Table~\ref{EXAMPLE2} gives hypothetical $p$-values and decisions;
calculations use exact allocations, with displayed levels rounded.

\paragraph{Retest.}
In Stage~1, the thresholds in $F_1$ and $F_2$ are $0.00625$
and $0.0041667$, producing no rejections. The threshold in
$F_3$ is $0.0020833$, rejecting $H_{32}$ with $p_{32}=0.0013$.
No-retest ends with this single rejection.
In Stage~2,
\[
\begin{aligned}
\alpha_{1(2)}&=\alpha_1+\tfrac14\alpha_3
\approx0.0135417,\\
\alpha_{2(2)}&=\alpha_2+\tfrac14\alpha_3=0.009375,\\
\alpha_{3(2)}&=\alpha_3+\tfrac14\alpha_{2(2)}
\approx0.0065104.
\end{aligned}
\]
No primary hypothesis is rejected, but $H_{22}$ satisfies
$0.0046\leq0.009375/2=0.0046875$.
There is no additional rejection in $F_3$.
In Stage~3,
\[
\alpha_{1(3)}=\alpha_1+\tfrac14\alpha_2
+\tfrac14\alpha_3=0.015625.
\]
Neither primary $p$-value meets threshold $0.0078125$;
the other levels and rejection sets are unchanged.
Retest terminates with $\{H_{22},H_{32}\}$.

\paragraph{BR and MR.}
Both start in $F_1$ at level $0.025$. Its smallest $p$-value,
$0.0200$, exceeds the first Bonferroni or truncated Holm
threshold $0.0125$. Neither proceeds to later families,
and neither makes a rejection.

\paragraph{SC.}
Initially only $H_{32}$ is rejected. The error fraction in
$F_3$ is $0.5+0.5/2=0.75$, so each other family receives
$0.5(1-0.75)\alpha_3=\alpha_3/8$. Updated levels are
\[
a_1=\alpha_1+\alpha_3/8\approx0.0130208,
\qquad a_2=\alpha_2+\alpha_3/8\approx0.0088542.
\]
Their first truncated Holm thresholds remain $a_1/2$ and
$a_2/2$, even as truncation parameters change. Since
$0.0200>0.0065104$ and $0.0046>0.0044271$, no further
rejection occurs. SC ends with only $H_{32}$.

\paragraph{Pooled Holm.}
The first two ordered $p$-values satisfy
$0.0013\leq0.025/6$ and $0.0046\leq0.025/5$.
The third fails because $0.0070>0.025/4=0.00625$.
Holm therefore rejects $H_{32}$ and $H_{22}$.

\begin{table}[htbp]
\centering
\small
\caption{Rejection decisions for a hypothetical
three-family example. Retest, No-retest, and SC use
initial allocations $\alpha/2$, $\alpha/3$, and
$\alpha/6$, with $\alpha=0.025$. Original BR and MR
initially assign the full level $\alpha$ to $F_1$.
Pooled Holm tests all six hypotheses
at the same global level.
S = significant; NS = not significant.}
\label{EXAMPLE2}
\begin{tabular}{llccccccc}
\hline
Family & Hypothesis & $p$-value & Retest & No-retest
& BR & MR & SC & Holm\\
\hline
$F_1$ & $H_{11}$ & 0.0200 & NS & NS & NS & NS & NS & NS\\
      & $H_{12}$ & 0.0240 & NS & NS & NS & NS & NS & NS\\
$F_2$ & $H_{21}$ & 0.0070 & NS & NS & NS & NS & NS & NS\\
      & $H_{22}$ & 0.0046 & S  & NS & NS & NS & NS & S\\
$F_3$ & $H_{31}$ & 0.0300 & NS & NS & NS & NS & NS & NS\\
      & $H_{32}$ & 0.0013 & S  & S  & NS & NS & S  & S\\
\hline
\multicolumn{3}{l}{Number rejected}
& 2 & 1 & 0 & 0 & 1 & 2\\
\hline
\end{tabular}
\end{table}

\paragraph{Interpretation.}
Partial rejection in $F_3$ allows Retest to reject $H_{22}$
through backward feedback. No-retest and SC reject only
$H_{32}$, while pooled Holm reaches Retest's rejection set.
BR and MR make no rejections because their initial gate fails;
their comparison also reflects restricted access to later
families, which receive positive initial allocations under
Retest, No-retest, and SC.
\section{A Two-Layer Extension}\label{sec:extension}

We extend the construction to two layers containing
multiple families. For example, each family may concern
an endpoint across several patient populations, with
families grouped into primary and secondary layers.
Families within a layer are tested simultaneously,
while the layers are tested sequentially within each
stage. The extension retains fixed Bonferroni component
tests and permits rejection-dependent transfers and
re-testing across layers.

\subsection*{Setup}

Let $F_{ij}$ denote family $j$ in layer $i$, where
$i=1,2$ and $j=1,\ldots,m_i$. Each family contains
$n_{ij}\geq1$ hypotheses, with
\[
m_1+m_2=m,
\qquad
\sum_{i=1}^{2}\sum_{j=1}^{m_i}n_{ij}=n.
\]
Assign initial levels $\alpha_{ij}\geq0$ satisfying
\[
\sum_{i=1}^{2}\sum_{j=1}^{m_i}\alpha_{ij}=\alpha.
\]

\begin{figure}[htbp]
\centering
\includegraphics[width=0.6\textwidth]{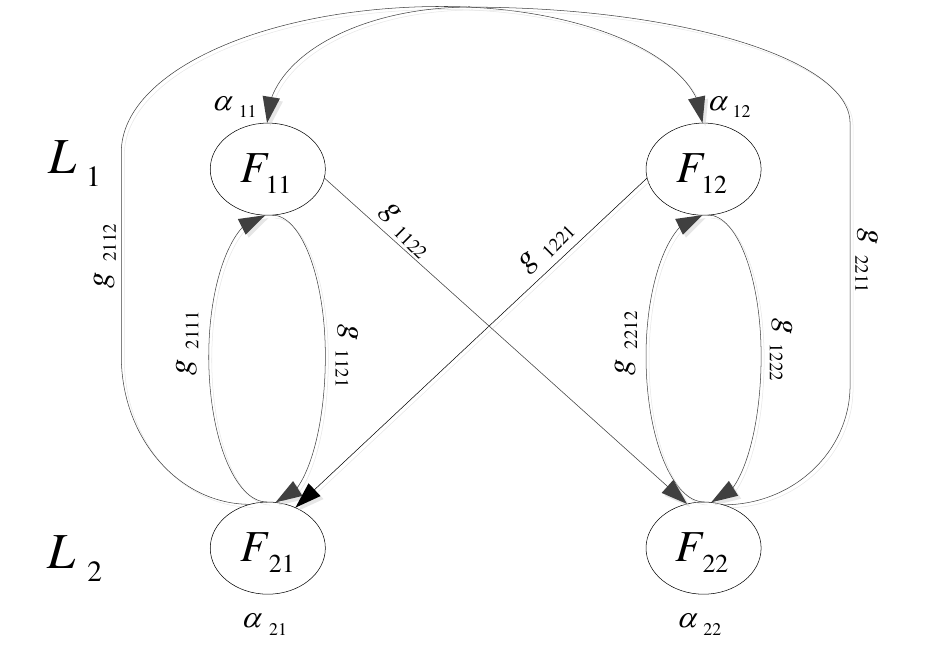}
\caption{Two-layer family-level graph with four families
and inter-layer transfers.}
\label{fig:twolayer}
\end{figure}

Let $g_{ijkl}$ specify the transition coefficient
from $F_{ij}$ to $F_{kl}$. Only inter-layer transfers
are permitted:
\[
g_{ijkl}\geq0,
\qquad
g_{ijkl}=0\quad\text{if }i=k,
\]
\[
\sum_{l=1}^{m_2}g_{1j2l}\leq1,
\qquad
\sum_{j=1}^{m_1}g_{2l1j}\leq1.
\]
Any remaining fraction is not transferred.
In the graphical representation, nodes correspond
to families and directed edges specify these
coefficients. Figure~\ref{fig:twolayer} illustrates
a four-family structure.

\subsection*{Procedure}

Write $\alpha_{ij(k)}$ for the testing level of $F_{ij}$
in stage $k$ and $R_{ij(k)}$ for its cumulative rejection
set. Bonferroni testing uses thresholds
$\alpha_{ij(k)}/n_{ij}$, retaining the original family
sizes and all earlier rejections.

\begin{algorithm}[htbp]
\caption{Two-layer Bonferroni procedure with iterative retesting}
\label{alg:twolayer}
\begin{flushleft}

\textbf{Initialization.}
Set $R_{ij(0)}=\varnothing$ for all families and $k=1$.

\medskip
\textbf{First stage.}
Test all layer-1 families simultaneously at their
initial levels $\alpha_{1j}$, obtaining $R_{1j(1)}$.
For each layer-2 family, compute
\[
\alpha_{2l(1)}
=\alpha_{2l}
+\sum_{j=1}^{m_1}
\frac{|R_{1j(1)}|}{n_{1j}}
g_{1j2l}\alpha_{1j}.
\]
Test the layer-2 families simultaneously at these
levels, obtaining $R_{2l(1)}$.

\medskip
\textbf{Subsequent stages.}
For $k\geq2$, first compute the layer-1 levels
\[
\alpha_{1j(k)}
=\alpha_{1j}
+\sum_{l=1}^{m_2}
\frac{|R_{2l(k-1)}|}{n_{2l}}
g_{2l1j}\alpha_{2l}.
\]
Retest the remaining hypotheses in all layer-1
families simultaneously, obtaining $R_{1j(k)}$.
Next compute
\[
\alpha_{2l(k)}
=\alpha_{2l}
+\sum_{j=1}^{m_1}
\frac{|R_{1j(k)}|}{n_{1j}}
g_{1j2l}\alpha_{1j(k)}.
\]
Retest the remaining hypotheses in all layer-2
families simultaneously, obtaining $R_{2l(k)}$.

\medskip
\textbf{Stopping rule.}
Stop if every rejection set is unchanged from stage
$k-1$, or all hypotheses have been rejected.
Otherwise increase $k$ by one and continue.

\end{flushleft}
\end{algorithm}

Forward contributions use current-stage layer-1 levels,
whereas backward contributions use initial layer-2
allocations and previous-stage rejection fractions.
Levels are recomputed at each stage rather than
accumulated repeatedly. They are nondecreasing,
rejection sets are nested, and the procedure terminates
after at most $n$ stages that add rejections and one
possible final stage without a new rejection.

\begin{remark}
\normalfont
With one family per layer and unit transitions in
both directions, Algorithm~\ref{alg:twolayer} reduces
to Algorithm~\ref{alg:twofamily}.
The inter-layer graph can also represent one-to-one,
one-to-many, or many-to-one transfer structures.
These connections specify contributions rather than
automatically imposing eligibility conditions.
Strict gatekeeping requirements must be encoded through
the initial allocations and permitted transfers, as
discussed in Section~\ref{sec:parameter-guidance}.
\end{remark}

\begin{theorem}\label{thm:twolayer}
Under valid marginal $p$-values and the allocation
and transition conditions above,
Algorithm~\ref{alg:twolayer} strongly controls
the global FWER at level $\alpha$ under arbitrary
dependence.
\end{theorem}

The proof is given in the Appendix and follows the
strategy of Theorem~\ref{thm:gfwer}: deterministic
bounds for the rejection-dependent testing levels
are combined with Bonferroni inequalities.
The guarantee concerns the probability of rejecting
a true null across all families; it does not require
the updated family levels to sum to at most $\alpha$.
The extension broadens the family structure while
retaining the Bonferroni component-test scope.
\section{Concluding Remarks}\label{sec:conclusion}

This paper develops a family-level graphical procedure
at the intersection of graphical multiple testing and
gatekeeping methods. The construction combines
partial-rejection-based transfers with sequential
re-testing, using fixed Bonferroni component tests and
a prespecified family transition matrix. Its updating
rule combines current-stage forward contributions based
on current family levels with previous-stage backward
contributions based on initial allocations. We establish
strong global FWER control at level $\alpha$ under
arbitrary dependence.

The contribution is an explicit family-level allocation
and updating scheme with a corresponding validity
guarantee. It permits feedback following partial rejection,
whereas Algorithm~1 of \citet{maurer2014note} requires
complete rejection before inter-family transfer.
It retains fixed component tests and transition
coefficients, whereas the specified superchain algorithm
uses component updates and graph reweighting
\citep{kordzakhia2013superchain}.
Relative to the forward-only framework of
\citet{qiu2026graphical}, it permits backward contributions
and repeated testing, with a narrower scope of component
tests. These distinctions position the procedure within
existing graphical and gatekeeping methods.

The numerical study demonstrates gains over No-retest
under identical initial allocations, component tests,
and first-pass transfer rules. Comparisons with the
original BR and MR procedures, SC, and pooled Holm
depend on the configuration. The matched hypothesis-level
graphical comparator contains Retest's rejection set
in every simulated replication, identifying an efficiency
cost of the proposed construction in the studied settings.
This empirical finding does not establish a general
dominance theorem. The clinical illustrations show how
partial transfers and backward feedback create additional
testing opportunities within the specified family-level
strategy.

For applications, initial allocations and permitted
transfers should reflect scientific priorities and
study-success criteria. Positive initial allocations
to later families allow testing without upstream
rejections, so strict gatekeeping restrictions must
be explicitly encoded. The family-level graph makes
these choices transparent, while the theoretical
guarantee addresses error control across all families.

Future work includes extensions to other component
tests, supported by suitable transfer rules and global
error-control arguments. Further research may develop
methods for choosing allocations and transition
coefficients according to prespecified study objectives,
and extend the construction to more complex hierarchies
and adaptive designs.
\section*{Appendix}

\setcounter{subsection}{0}
\renewcommand{\thesubsection}{A.\arabic{subsection}}

\counterwithin{lemma}{subsection}
\setcounter{lemma}{0}
\renewcommand{\thelemma}{\thesubsection.\arabic{lemma}}

\subsection{Proof of Proposition~\ref{prop:twofamily}}
\label{app:proof-twofamily}

\begin{proof}
We prove the three statements in turn.

\medskip
\noindent
\textit{(i) Monotonicity of the local significance levels.}
Recall that
\[
\alpha_{1(1)}=\alpha_1,\qquad
\alpha_{2(1)}=\alpha_2+\frac{|R_{1(1)}|}{n_1}\alpha_1,
\]
and for each $k\ge2$,
\begin{align}
\alpha_{1(k)} &= \alpha_1+\frac{|R_{2(k-1)}|}{n_2}\alpha_2, \label{eq:app-a1k}\\
\alpha_{2(k)} &= \alpha_2+\frac{|R_{1(k)}|}{n_1}\alpha_{1(k)}. \label{eq:app-a2k}
\end{align}

We first show that the rejection sets are nondecreasing in $k$, that is,
\[
R_{i(k-1)}\subseteq R_{i(k)}, \qquad i=1,2,\quad k\ge2.
\]
Indeed, the Bonferroni procedure rejects hypothesis $H_{ij}\in F_i$ at stage $k$ whenever
\[
p_{ij}\le \frac{\alpha_{i(k)}}{n_i}.
\]
Thus, if $\alpha_{i(k)}\ge \alpha_{i(k-1)}$, then every hypothesis rejected at stage $k-1$ remains rejected at stage $k$. Hence, to establish the monotonicity of the rejection sets, it suffices to verify that the sequences $\{\alpha_{1(k)}\}_{k\ge1}$ and $\{\alpha_{2(k)}\}_{k\ge1}$ are nondecreasing.

We prove this by induction on $k$.

First, from \eqref{eq:app-a1k},
\[
\alpha_{1(2)}
=
\alpha_1+\frac{|R_{2(1)}|}{n_2}\alpha_2
\ge \alpha_1
=
\alpha_{1(1)}.
\]
Also,
\[
\alpha_{2(2)}
=
\alpha_2+\frac{|R_{1(2)}|}{n_1}\alpha_{1(2)}
\ge
\alpha_2+\frac{|R_{1(1)}|}{n_1}\alpha_1
=
\alpha_{2(1)},
\]
because $|R_{1(2)}|\ge |R_{1(1)}|$ and $\alpha_{1(2)}\ge \alpha_{1(1)}$.

Now assume for some $k\ge2$ that
\[
\alpha_{1(k)}\ge \alpha_{1(k-1)}
\quad\text{and}\quad
\alpha_{2(k)}\ge \alpha_{2(k-1)}.
\]
Then, by the Bonferroni rejection rule, we have
\[
R_{1(k-1)}\subseteq R_{1(k)}
\quad\text{and}\quad
R_{2(k-1)}\subseteq R_{2(k)}.
\]
Therefore,
\[
|R_{2(k)}|\ge |R_{2(k-1)}|.
\]
Using \eqref{eq:app-a1k}, it follows that
\[
\alpha_{1(k+1)}
=
\alpha_1+\frac{|R_{2(k)}|}{n_2}\alpha_2
\ge
\alpha_1+\frac{|R_{2(k-1)}|}{n_2}\alpha_2
=
\alpha_{1(k)}.
\]
Similarly, since $|R_{1(k+1)}|\ge |R_{1(k)}|$ and $\alpha_{1(k+1)}\ge \alpha_{1(k)}$, \eqref{eq:app-a2k} yields
\[
\alpha_{2(k+1)}
=
\alpha_2+\frac{|R_{1(k+1)}|}{n_1}\alpha_{1(k+1)}
\ge
\alpha_2+\frac{|R_{1(k)}|}{n_1}\alpha_{1(k)}
=
\alpha_{2(k)}.
\]
This completes the induction. Hence both sequences $\{\alpha_{1(k)}\}_{k\ge1}$ and $\{\alpha_{2(k)}\}_{k\ge1}$ are nondecreasing.

\medskip
\noindent
\textit{(ii) Nestedness of the rejection sets.}
Fix $i\in\{1,2\}$ and $k\ge2$. Since the Bonferroni procedure for family $F_i$ at stage $k$ rejects exactly those hypotheses with
\[
p_{ij}\le \frac{\alpha_{i(k)}}{n_i},
\]
and since part (i) shows that $\alpha_{i(k)}\ge \alpha_{i(k-1)}$, we have
\[
\frac{\alpha_{i(k)}}{n_i}\ge \frac{\alpha_{i(k-1)}}{n_i}.
\]
Therefore every hypothesis rejected at stage $k-1$ is also rejected at stage $k$, so
\[
R_{i(k-1)}\subseteq R_{i(k)}.
\]
By induction over $k$, this gives
\[
R_{i(1)}\subseteq R_{i(2)}\subseteq \cdots, \qquad i=1,2.
\]

\medskip
\noindent
\textit{(iii) Finite-step termination.}
By part (ii), the rejection sets are nested. Hence, as $k$ increases, each family can only gain new rejections and can never lose previously rejected hypotheses. Since family $F_i$ contains only $n_i$ hypotheses, the number of times $R_{i(k)}$ can strictly increase is at most $n_i$.

The algorithm stops once no new rejection occurs at a stage. Therefore, each nonterminal stage must produce at least one new rejection in either $F_1$ or $F_2$. Since the total number of hypotheses across both families is $n_1+n_2$, there can be at most $n_1+n_2$ such stages. Consequently, the procedure terminates after finitely many iterations, with at most $n_1+n_2$ stages adding rejections and one
possible final stage without a new rejection, for a total
bound of $n_1+n_2+1$ stages.

This proves the proposition.
\end{proof}

\subsection{Proof of Theorem~\ref{thm:gfwer}}

\begin{proof}
The proof proceeds by first controlling the familywise error rate up to an arbitrary stage $k$, and then passing to the limit as $k\to\infty$.

For each $k\ge 1$, let $V_k$ denote the total number of false rejections across all families up to and including stage $k$, and define
\[
\operatorname{FWER}_k = \Pr(V_k \ge 1).
\]
Then the global familywise error rate is
\[
\operatorname{FWER}
=
\Pr\Bigl(\bigcup_{k=1}^{\infty}\{V_k \ge 1\}\Bigr).
\]

For each stage $j\ge 1$, let $D_j$ denote the event that the
cumulative rejection sets at stage $j$ contain a true null, and let
\[
E_{i(j)}
=
\{R_{i(j)}\text{ contains at least one true null hypothesis}\}.
\]
Then
\[
D_j = \bigcup_{i=1}^m E_{i(j)},
\qquad
\{V_k \ge 1\} = \bigcup_{j=1}^k D_j,
\]
so that
\[
\operatorname{FWER}_k
=
\Pr\Bigl(\bigcup_{j=1}^k D_j\Bigr).
\]

Let $D_k$ denote the event that the cumulative rejection sets
at stage $k$ contain a true null, and let
$\overline{D}_k=D_k^c$ denote its complement, the event
that no true null hypothesis is rejected at that stage.

We first show that
\begin{equation}\label{eq:A1-reduction}
\operatorname{FWER}_k
=1-\Pr(\overline{D}_k).
\end{equation}
The local critical values are nondecreasing across stages,
so the rejection sets are nested. Consequently,
$\overline{D}_k\subseteq\overline{D}_j$ for every $j<k$:
if no true null hypothesis is rejected at stage $k$,
none was rejected at an earlier stage. Hence,
\[
1-\operatorname{FWER}_k
=
\Pr\Bigl(\bigcap_{j=1}^{k}\overline{D}_j\Bigr)
=
\Pr(\overline{D}_k),
\]
which proves \eqref{eq:A1-reduction}.

It therefore suffices to show that
\[
1-\Pr(\overline{D}_k)\le\alpha
\qquad\text{for every }k\ge1.
\]

Now let $T_i$ denote the set of true null hypotheses in family $F_i$, and write
\[
t_i=\frac{|T_i|}{n_i},
\qquad
q_i=1-\frac{|T_i|}{n_i},
\qquad i=1,\dots,m.
\]
On the event $\overline{D}_k$, no true null hypothesis is rejected at stage $k$. By the monotonicity argument above, no true null hypothesis is rejected at any of the first $k$ stages. Hence, for every family $F_i$,
\[
|R_{i(s)}|
\le n_i-|T_i|
=
q_i n_i,
\qquad s=1,\dots,k.
\]
Comparing this with the updating rule in Algorithm~\ref{alg:multifamily}, we obtain
\[
\alpha_{i(k)} \le \alpha_{i(k)}^*,
\qquad i=1,\dots,m,
\]
where $\alpha_{i(k)}^*$ is defined recursively by
\begin{align}
\alpha_{1(k)}^*
&=
\alpha_1+\sum_{l=2}^m q_l g_{l1}\alpha_l, \label{eq:A1-alpha1star}\\
\alpha_{i(k)}^*
&=
\alpha_i
+\sum_{j=1}^{i-1} q_j g_{ji}\alpha_{j(k)}^*
+\sum_{l=i+1}^{m} q_l g_{li}\alpha_l,
\qquad i=2,\dots,m. \label{eq:A1-alphaistar}
\end{align}
To justify the probability comparison below, consider the first
family test producing a false rejection, ordered by stage and
then by family. Before this test, every rejection fraction is
at most its corresponding $q_i$. Induction over the preceding
family tests therefore bounds their levels, and the level of
the test producing the first false rejection, by the deterministic
quantities above. These bounds do not depend on the stage index;
the absence of backward terms in stage~1 only lowers its levels.
Thus, if all true-null $p$-values exceed these deterministic
thresholds, a first false rejection cannot occur. Consequently,
\begin{align}
\Pr(\overline{D}_k)
&\ge
\Pr\Bigl(
\bigcap_{i=1}^m\bigcap_{H_{ij}\in T_i}
\Bigl\{p_{ij}>\frac{\alpha_{i(k)}^*}{n_i}\Bigr\}
\Bigr).
\label{eq:A1-keybound}
\end{align}

Applying the union bound to \eqref{eq:A1-keybound} and using the super-uniformity of true-null $p$-values, we get
\begin{align}
1-\Pr(\overline{D}_k)
&\le
\sum_{i=1}^m
\Pr\Bigl(
\bigcup_{H_{ij}\in T_i}
\Bigl\{p_{ij}\le \frac{\alpha_{i(k)}^*}{n_i}\Bigr\}
\Bigr) \nonumber\\
&\le
\sum_{i=1}^m \frac{|T_i|}{n_i}\alpha_{i(k)}^*
=
\sum_{i=1}^m t_i \alpha_{i(k)}^*. \label{eq:A1-union}
\end{align}

It remains to show that
\begin{equation}\label{eq:A1-target}
\sum_{i=1}^m t_i \alpha_{i(k)}^* \le \alpha.
\end{equation}

Define, for $j=1,\dots,m-1$,
\begin{align}
f(j)
&=
\sum_{i=1}^{j}
\left[
t_i+q_i\Bigl(\sum_{l=j+1}^{m} g_{il}\Bigr)
\right]\alpha_{i(k)}^*
+
\sum_{l=j+1}^{m}
\left[
t_l+q_l\Bigl(\sum_{i=j+1}^{m} g_{li}\Bigr)
\right]\alpha_l.
\label{eq:A1-fj}
\end{align}
By Lemma~\ref{lem:monotone},
\[
f(m-1)\le f(m-2)\le \cdots \le f(1).
\]

Now, using \eqref{eq:A1-alphaistar} with $i=m$, we have
\[
\alpha_{m(k)}^*
=
\alpha_m+\sum_{i=1}^{m-1} q_i g_{im}\alpha_{i(k)}^*.
\]
Therefore,
\begin{align}
\sum_{i=1}^m t_i\alpha_{i(k)}^*
&=
\sum_{i=1}^{m-1} t_i\alpha_{i(k)}^*
+
t_m\Bigl(\alpha_m+\sum_{i=1}^{m-1} q_i g_{im}\alpha_{i(k)}^*\Bigr) \nonumber\\
&=
\sum_{i=1}^{m-1}\bigl[t_i+t_m q_i g_{im}\bigr]\alpha_{i(k)}^*
+t_m\alpha_m.
\end{align}

At this point, note that
\[
t_m q_i g_{im}\le q_i g_{im},
\]
since $0\le t_m\le 1$. Hence
\begin{align}
\sum_{i=1}^m t_i\alpha_{i(k)}^*
&\le
\sum_{i=1}^{m-1}\bigl[t_i+ q_i g_{im}\bigr]\alpha_{i(k)}^*
+t_m\alpha_m
=
f(m-1).
\label{eq:A1-fm1}
\end{align}
Therefore,
\[
\sum_{i=1}^m t_i\alpha_{i(k)}^* \le f(m-1)\le f(1).
\]

It remains to evaluate $f(1)$. By definition,
\[
f(1)=
\left[t_1+q_1\sum_{l=2}^m g_{1l}\right]\alpha_{1(k)}^*
+
\sum_{l=2}^{m}
\left[t_l+q_l\sum_{i=2}^{m} g_{li}\right]\alpha_l.
\]
Since $\sum_{l=1}^m g_{1l}\le1$ and $g_{11}=0$, we have
$\sum_{l=2}^m g_{1l}\le1$, and thus
\[
t_1+q_1\sum_{l=2}^m g_{1l}\le t_1+q_1=1.
\]
Also, since $\sum_{i=1}^m g_{li}\le1$ and $g_{ll}=0$,
\[
\sum_{i=2}^m g_{li}\le1-g_{l1},
\qquad l=2,\dots,m.
\]
Therefore, using \eqref{eq:A1-alpha1star},
\begin{align}
f(1)
&\le
\alpha_{1(k)}^*
+
\sum_{l=2}^{m}\bigl[t_l+q_l(1-g_{l1})\bigr]\alpha_l \nonumber\\
&=
\alpha_{1(k)}^*
+
\sum_{l=2}^{m}(1-q_l g_{l1})\alpha_l \nonumber\\
&=
\left(\alpha_1+\sum_{l=2}^m q_l g_{l1}\alpha_l\right)
+
\sum_{l=2}^{m}(1-q_l g_{l1})\alpha_l \nonumber\\
&=
\sum_{l=1}^{m}\alpha_l
=
\alpha. \label{eq:A1-f1}
\end{align}
Combining \eqref{eq:A1-fm1} and \eqref{eq:A1-f1}, we obtain \eqref{eq:A1-target}. Hence, by \eqref{eq:A1-union},
\[
1-\Pr(\overline{D}_k)\le \alpha.
\]
Together with \eqref{eq:A1-reduction}, this yields
\[
\operatorname{FWER}_k\le \alpha
\qquad \text{for every } k\ge 1.
\]

Finally, since $V_k$ is nondecreasing in $k$, the events $\{V_k\ge 1\}$ form an increasing sequence. Therefore,
\[
\operatorname{FWER}
=
\Pr\Bigl(\bigcup_{k=1}^{\infty}\{V_k\ge 1\}\Bigr)
=
\lim_{k\to\infty}\Pr(V_k\ge 1)
=
\lim_{k\to\infty}\operatorname{FWER}_k
\le \alpha.
\]
This completes the proof.
\end{proof}

\subsection{A monotonicity lemma for level allocation}

The following lemma is the key ingredient used in the proof above.

\begin{lemma}\label{lem:monotone}
With $t_i=|T_i|/n_i$ and $q_i=1-|T_i|/n_i$, define, for $j=1,\dots,m-1$,
\begin{align}
f(j)
&=
\sum_{i=1}^{j}
\left[
t_i+q_i\Bigl(\sum_{l=j+1}^{m} g_{il}\Bigr)
\right]\alpha_{i(k)}^*
+
\sum_{l=j+1}^{m}
\left[
t_l+q_l\Bigl(\sum_{i=j+1}^{m} g_{li}\Bigr)
\right]\alpha_l.
\label{eq:A2-fj}
\end{align}
Then $f(j)$ is nonincreasing in $j$, that is,
\[
f(j)\le f(j-1),
\qquad j=2,\dots,m-1.
\]
\end{lemma}

\begin{proof}
Fix $j\in\{2,\dots,m-1\}$. Write
\[
c_j=t_j+q_j\sum_{l=j+1}^{m}g_{jl}.
\]
Since $0\le t_j\le1$, $0\le q_j\le1$, and $\sum_{l=j+1}^{m}g_{jl}\le1$, we have
\[
0\le c_j\le t_j+q_j =1.
\]

Now expand the definition of $f(j)$ by isolating the term corresponding to $i=j$:
\begin{align}
f(j)
&=
\sum_{i=1}^{j-1}
\left[
t_i+q_i\Bigl(\sum_{l=j+1}^{m} g_{il}\Bigr)
\right]\alpha_{i(k)}^*
+
c_j\,\alpha_{j(k)}^* \nonumber\\
&\quad
+
\sum_{l=j+1}^{m}
\left[
t_l+q_l\Bigl(\sum_{i=j+1}^{m} g_{li}\Bigr)
\right]\alpha_l.
\label{eq:A2-expand}
\end{align}

From the recursive definition of $\alpha_{j(k)}^*$,
\[
\alpha_{j(k)}^*
=
\alpha_j
+
\sum_{i=1}^{j-1} q_i g_{ij}\alpha_{i(k)}^*
+
\sum_{l=j+1}^{m} q_l g_{lj}\alpha_l.
\]
Multiplying both sides by $c_j$ and using $0\le c_j\le1$, we obtain
\[
c_j\alpha_{j(k)}^*
\le
c_j\alpha_j
+
\sum_{i=1}^{j-1} q_i g_{ij}\alpha_{i(k)}^*
+
\sum_{l=j+1}^{m} q_l g_{lj}\alpha_l.
\]
Substituting this bound into \eqref{eq:A2-expand} yields
\begin{align}
f(j)
&\le
\sum_{i=1}^{j-1}
\left[
t_i+q_i\Bigl(\sum_{l=j+1}^{m} g_{il}\Bigr)
\right]\alpha_{i(k)}^*
+
\sum_{i=1}^{j-1} q_i g_{ij}\alpha_{i(k)}^*
\nonumber\\
&\quad
+
c_j\alpha_j
+
\sum_{l=j+1}^{m}
\left[
t_l+q_l\Bigl(\sum_{i=j+1}^{m} g_{li}\Bigr)
\right]\alpha_l
+
\sum_{l=j+1}^{m} q_l g_{lj}\alpha_l.
\label{eq:A2-sub}
\end{align}

We now regroup terms. For $i=1,\dots,j-1$,
\[
\left[
t_i+q_i\Bigl(\sum_{l=j+1}^{m} g_{il}\Bigr)
\right]
+
q_i g_{ij}
=
t_i+q_i\Bigl(\sum_{l=j}^{m} g_{il}\Bigr).
\]
Also, since
\[
c_j=t_j+q_j\sum_{l=j+1}^{m}g_{jl}
=
t_j+q_j\sum_{i=j+1}^{m}g_{ji},
\]
the term $c_j\alpha_j$ is exactly
\[
\left[
t_j+q_j\Bigl(\sum_{i=j+1}^{m}g_{ji}\Bigr)
\right]\alpha_j.
\]
For $l=j+1,\dots,m$,
\[
\left[
t_l+q_l\Bigl(\sum_{i=j+1}^{m} g_{li}\Bigr)
\right]
+
q_l g_{lj}
=
t_l+q_l\Bigl(\sum_{i=j}^{m} g_{li}\Bigr).
\]

Substituting these identities into \eqref{eq:A2-sub}, we obtain
\begin{align}
f(j)
&\le
\sum_{i=1}^{j-1}
\left[
t_i+q_i\Bigl(\sum_{l=j}^{m} g_{il}\Bigr)
\right]\alpha_{i(k)}^*
+
\left[
t_j+q_j\Bigl(\sum_{i=j+1}^{m} g_{ji}\Bigr)
\right]\alpha_j
\nonumber\\
&\quad
+
\sum_{l=j+1}^{m}
\left[
t_l+q_l\Bigl(\sum_{i=j}^{m} g_{li}\Bigr)
\right]\alpha_l.
\end{align}
This is exactly $f(j-1)$. Therefore,
\[
f(j)\le f(j-1),
\qquad j=2,\dots,m-1.
\]
Hence $f(j)$ is nonincreasing in $j$.
\end{proof}

\subsection{ Proof of Theorem~\ref{thm:twolayer}}

\begin{proof}
For each $k \ge 1$, let $V_k$ denote the total number of false rejections across all families up to and including stage $k$, and define
\[
\operatorname{FWER}_k=\Pr(V_k\ge 1).
\]
Then the global familywise error rate is
\[
\operatorname{FWER}
=
\Pr\Bigl(\bigcup_{k=1}^{\infty}\{V_k\ge 1\}\Bigr).
\]

For each stage $j\ge1$, let $D_j$ denote the event that the
cumulative rejection sets at stage $j$ contain a true null,
and put $\overline D_j=D_j^c$. Then
\[
\{V_k\ge 1\}=\bigcup_{j=1}^k D_j,
\qquad
\operatorname{FWER}_k=\Pr\Bigl(\bigcup_{j=1}^k D_j\Bigr).
\]

As in the proof of Theorem~\ref{thm:gfwer}, the local critical values are nondecreasing across stages. Hence, if no true null hypothesis is rejected at stage $k$, then no true null hypothesis can have been rejected at any earlier stage. Therefore,
\[
1-\operatorname{FWER}_k=\Pr(\overline{D}_k),
\]
or equivalently,
\begin{equation}\label{eq:twolayer_reduction}
\operatorname{FWER}_k = 1-\Pr(\overline{D}_k).
\end{equation}
Thus, it is enough to show that
\[
1-\Pr(\overline{D}_k)\le \alpha
\qquad \text{for every } k\ge 1.
\]

For $i=1,2$, let $F_{ij}$, $j=1,\dots,m_i$, denote the families in layer $i$. For each hypothesis $H_{ijs}$ in family $F_{ij}$, let $p_{ijs}$ be its $p$-value, where $s=1,\dots,n_{ij}$. Let $T_{ij}$ denote the set of true null hypotheses in $F_{ij}$, and write
\[
t_{ij}=\frac{|T_{ij}|}{n_{ij}},
\qquad
q_{ij}=1-\frac{|T_{ij}|}{n_{ij}}.
\]

On the event $\overline{D}_k$, no true null hypothesis is rejected at stage $k$, and hence, by monotonicity, none has been rejected at any earlier stage. Therefore, for every family $F_{ij}$,
\[
|R_{ij(s)}|\le n_{ij}-|T_{ij}| = q_{ij}n_{ij},
\qquad s=1,\dots,k.
\]
Comparing this with the updating rule in Algorithm~\ref{alg:twolayer}, we obtain
\[
\alpha_{1j(k)}\le \alpha_{1j(k)}^*,
\qquad
\alpha_{2l(k)}\le \alpha_{2l(k)}^*,
\]
where
\begin{align}
\alpha_{1j(k)}^*
&=
\alpha_{1j}+\sum_{l=1}^{m_2} q_{2l}\, g_{2l1j}\,\alpha_{2l},
\qquad j=1,\dots,m_1, \label{eq:twolayer_star1}\\
\alpha_{2l(k)}^*
&=
\alpha_{2l}+\sum_{j=1}^{m_1} q_{1j}\, g_{1j2l}\,\alpha_{1j(k)}^*,
\qquad l=1,\dots,m_2. \label{eq:twolayer_star2}
\end{align}

As in the preceding proof, the deterministic bounds apply to
the family test producing the first false rejection, because
all rejection fractions used before that test are bounded by
their false-null proportions. Families within a layer can be
ordered arbitrarily for this argument, since there are no
within-layer transfers. Thus, exceeding all deterministic
true-null thresholds rules out a first false rejection.

Hence,
\begin{align}
\Pr(\overline{D}_k)
&\ge
\Pr\Biggl(
\bigcap_{j=1}^{m_1}\bigcap_{H_{1js}\in T_{1j}}
\Bigl\{p_{1js}>\frac{\alpha_{1j(k)}^*}{n_{1j}}\Bigr\}
\;\cap\;
\bigcap_{l=1}^{m_2}\bigcap_{H_{2ls}\in T_{2l}}
\Bigl\{p_{2ls}>\frac{\alpha_{2l(k)}^*}{n_{2l}}\Bigr\}
\Biggr).
\label{eq:twolayer_keybound}
\end{align}

Applying the union bound to \eqref{eq:twolayer_keybound} and using the super-uniformity of true-null $p$-values, we obtain
\begin{align}
1-\Pr(\overline{D}_k)
&\le
\sum_{j=1}^{m_1}
\Pr\Biggl(
\bigcup_{H_{1js}\in T_{1j}}
\Bigl\{p_{1js}\le \frac{\alpha_{1j(k)}^*}{n_{1j}}\Bigr\}
\Biggr)
\nonumber\\
&\quad
+
\sum_{l=1}^{m_2}
\Pr\Biggl(
\bigcup_{H_{2ls}\in T_{2l}}
\Bigl\{p_{2ls}\le \frac{\alpha_{2l(k)}^*}{n_{2l}}\Bigr\}
\Biggr)
\nonumber\\
&\le
\sum_{j=1}^{m_1} t_{1j}\alpha_{1j(k)}^*
+
\sum_{l=1}^{m_2} t_{2l}\alpha_{2l(k)}^*.
\label{eq:twolayer_union}
\end{align}

Now substitute \eqref{eq:twolayer_star2} into the second sum:
\begin{align}
\sum_{j=1}^{m_1} t_{1j}\alpha_{1j(k)}^*
+
\sum_{l=1}^{m_2} t_{2l}\alpha_{2l(k)}^*
&=
\sum_{j=1}^{m_1} t_{1j}\alpha_{1j(k)}^*
+
\sum_{l=1}^{m_2} t_{2l}
\left(
\alpha_{2l}
+
\sum_{j=1}^{m_1} q_{1j} g_{1j2l}\alpha_{1j(k)}^*
\right)
\nonumber\\
&=
\sum_{j=1}^{m_1}
\left(
t_{1j}
+
q_{1j}\sum_{l=1}^{m_2} t_{2l} g_{1j2l}
\right)\alpha_{1j(k)}^*
+
\sum_{l=1}^{m_2} t_{2l}\alpha_{2l}.
\label{eq:twolayer_expand}
\end{align}

Since $0\le t_{2l}\le 1$ for all $l$,
\[
\sum_{l=1}^{m_2} t_{2l} g_{1j2l}
\le
\sum_{l=1}^{m_2} g_{1j2l}
\le1.
\]
Therefore,
\[
t_{1j}
+
q_{1j}\sum_{l=1}^{m_2} t_{2l} g_{1j2l}
\le
t_{1j}+q_{1j}=1,
\]
and \eqref{eq:twolayer_expand} yields
\begin{align}
\sum_{j=1}^{m_1} t_{1j}\alpha_{1j(k)}^*
+
\sum_{l=1}^{m_2} t_{2l}\alpha_{2l(k)}^*
&\le
\sum_{j=1}^{m_1}\alpha_{1j(k)}^*
+
\sum_{l=1}^{m_2} t_{2l}\alpha_{2l}.
\label{eq:twolayer_mid}
\end{align}

Next, using \eqref{eq:twolayer_star1},
\begin{align}
\sum_{j=1}^{m_1}\alpha_{1j(k)}^*
+
\sum_{l=1}^{m_2} t_{2l}\alpha_{2l}
&=
\sum_{j=1}^{m_1}
\left(
\alpha_{1j}
+
\sum_{l=1}^{m_2} q_{2l} g_{2l1j}\alpha_{2l}
\right)
+
\sum_{l=1}^{m_2} t_{2l}\alpha_{2l}
\nonumber\\
&=
\sum_{j=1}^{m_1}\alpha_{1j}
+
\sum_{l=1}^{m_2}
q_{2l}
\left(\sum_{j=1}^{m_1} g_{2l1j}\right)\alpha_{2l}
+
\sum_{l=1}^{m_2} t_{2l}\alpha_{2l}.
\end{align}
Since the transition coefficients satisfy
\[
\sum_{j=1}^{m_1} g_{2l1j}\le1,
\qquad l=1,\dots,m_2,
\]
it follows that
\begin{align}
\sum_{j=1}^{m_1}\alpha_{1j(k)}^*
+
\sum_{l=1}^{m_2} t_{2l}\alpha_{2l}
&\le
\sum_{j=1}^{m_1}\alpha_{1j}
+
\sum_{l=1}^{m_2}(q_{2l}+t_{2l})\alpha_{2l}
\nonumber\\
&=
\sum_{j=1}^{m_1}\alpha_{1j}
+
\sum_{l=1}^{m_2}\alpha_{2l}
=
\alpha.
\label{eq:twolayer_finalbound}
\end{align}

Combining \eqref{eq:twolayer_union}, \eqref{eq:twolayer_mid}, and \eqref{eq:twolayer_finalbound}, we conclude that
\[
1-\Pr(\overline{D}_k)\le \alpha.
\]
Hence, by \eqref{eq:twolayer_reduction},
\[
\operatorname{FWER}_k\le \alpha
\qquad \text{for every } k\ge 1.
\]

Finally, since $V_k$ is nondecreasing in $k$, the events $\{V_k\ge 1\}$ form an increasing sequence. Therefore,
\[
\operatorname{FWER}
=
\Pr\Bigl(\bigcup_{k=1}^{\infty}\{V_k\ge 1\}\Bigr)
=
\lim_{k\to\infty}\Pr(V_k\ge 1)
=
\lim_{k\to\infty}\operatorname{FWER}_k
\le \alpha.
\]
This completes the proof.
\end{proof}

\bibliographystyle{plainnat}
\bibliography{ref_mcp}   

\end{document}